%% file: paper.tex
\documentclass[a4paper,UKenglish,cleveref, autoref]{oasics-v2019}

\usepackage[utf8]{inputenc}
\usepackage[ruled]{algorithm2e} 
\usepackage{todonotes}
\usepackage{numprint}
\usepackage{amsmath}

\newcommand{\dupper}{\overline{d}}
\newcommand{\Increment}{\raisebox{.12ex}{\hbox{\tt ++}}}

\newcommand{\mydist}{\mathrm{dist}}
\newcommand{\mystalldist}{\mathrm{stallDist}}

\DeclareUnicodeCharacter{3BC}{$\mu$}

\newcommand\footnoteref[1]{\protected@xdef\@thefnmark{\ref{#1}}\@footnotemark}
\usepackage{footnote}
\makesavenoteenv{tabular}
\makesavenoteenv{table}

\title{More Hierarchy in Route Planning Using Edge Hierarchies}

\author{Demian Hespe}{Karlsruhe Institute of Technology, Germany \and \url{www.kit.edu} }{hespe@kit.edu}{}{}

\author{Peter Sanders}{Karlsruhe Institute of Technology, Germany \and \url{www.kit.edu} }{sanders@kit.edu}{}{}

\authorrunning{D. Hespe and P. Sanders}

\Copyright{Demian Hespe and Peter Sanders}



\category{}

\relatedversion{}

\supplement{}

\acknowledgements{}

\nolinenumbers 

\hideOASIcs  

\EventEditors{Valentina Cacchiani and Alberto Marchetti-Spaccamela}
\EventNoEds{2}
\EventLongTitle{19th Symposium on Algorithmic Approaches for Transportation Modelling, Optimization, and Systems (ATMOS 2019)}
\EventShortTitle{ATMOS 2019}
\EventAcronym{ATMOS}
\EventYear{2019}
\EventDate{September 12--13, 2019}
\EventLocation{Munich, Germany}
\EventLogo{}
\SeriesVolume{75}
\ArticleNo{9}

\begin{document}

\maketitle

\begin{abstract}
A highly successful approach to route planning in networks
(particularly road networks) is to identify a hierarchy
in the network that allows faster queries after some preprocessing
that basically inserts additional ``shortcut''-edges into a graph.  In
the past there has been a succession of techniques that infer a more
and more fine grained hierarchy enabling increasingly more efficient
queries. This appeared to culminate in contraction hierarchies that assign
one hierarchy level to each \emph{vertex}.

In this paper we show how to identify an even more fine grained
hierarchy that assigns one level to each \emph{edge} of the
network. Our findings indicate that this can lead to considerably
smaller search spaces in terms of visited edges. Currently, this
rarely implies improved query times so that it remains an open question
whether edge hierarchies can lead to consistently improved
performance. However, we believe that the technique as such is a
noteworthy enrichment of the portfolio of available techniques that
might prove useful in the future.
\end{abstract}

\section{Introduction}
\label{sec:intro}

Computing shortest, fastest, or otherwise optimal routes in networks
is a fundamental problem needed to be solved in many applications. For road networks
alone there are multiple important applications, e.g., car navigation,
traffic simulation, planning in logistics, etc. An important
approach to fast route planning is to preprocess the network in such a
way that subsequent queries are accelerated. In this paper we focus on
point-to-point queries in road networks but note that other types of queries or networks might also
be supported in a way analogous to previous applications of
contraction hierarchies \cite{geisberger2012exact,bast2016route}.

A particularly successful class of preprocessing techniques for road
networks is to exploit hierarchy in the network. An informal way to
describe this is, that ``usually'', the farther away we are from
source or destination, the more important are the roads we use.
Hierarchical route planning techniques had a history in becoming more
aggressive in exploiting the hierarchy resulting in smaller
and smaller search spaces.  This began with early heuristics based on
road categories \cite{ishikawa1991map,jagadeesh2002heuristic} and
later used exact techniques that insert \emph{shortcut
  edges}. Shortcuts encode that certain subpaths are important and,
together with an appropriate query algorithm, ensure that optimal
paths can be found using hierarchical routing techniques. Such
techniques include overlay graphs \cite{SWZ02,delling2015customizable}, reach
based routing \cite{Gut04}, highway hierarchies
\cite{sanders2005highway} and highway node routing \cite{SS07a} -- so
far culminating in contraction hierarchies (CHs)~
\cite{GSSD08,geisberger2012exact,dibbelt2016customizable}.

CHs order the \emph{vertices} of the network by importance, i.e., we
conceptually have $n$ levels of hierarchy in a network with $n$
vertices. By inserting appropriate shortcuts, CHs ensure that there
exists an \emph{up-down path} between any pair of vertices that is a
shortest path.  An up-down path progresses from the source vertex to
more important vertices and then descends to less
important vertices until reaching the destination. CHs are widely used
since they are simple, allow fast preprocessing using little space and
lead to very small search space.

In this paper we introduce \emph{edge hierarchies} (EHs) as an even
more fine grained way to define hierarchy in the network.  EHs order
\emph{edges} rather than vertices by importance. They keep the concept of
up-down paths resulting in a very simple query algorithm.
Intuitively, this should further reduce search spaces. EHs -- in
contrast to CHs -- only have to explore edges out of a vertex $v$ that
are more important than the edge leading to $v$ in the current query.
Also note that EHs are very close to the informal definition of hierarchical routing
that we gave above.

After introducing basic terms and techniques in
Section~\ref{sec:prelim} and discussing further related work in
Section~\ref{sec:relatedwork}, we describe EHs in detail
in Section~\ref{sec:main}. While the basic query algorithm is simple
by design, a preprocessing algorithm finding the ``right'' shortcuts
turns out to be much more complicated.  We also discuss some basic
techniques for pruning the query search space.

In Section~\ref{sec:experiments} we perform an experimental evaluation
using large real world road networks and different cost functions.  It
turns out that EHs relax significantly less edges than CHs in
particular for cost functions that are known to be difficult for CHs --
with distance as the main optimization criterion and/or explicit
modeling of turn penalties. Unfortunately, the overall query time is usually
slightly worse than CHs and preprocessing time is considerably larger.
Overall, EHs are thus an intriguing concept with considerable
potential but they need further research to find out whether they will
eventually be useful in some applications.  In
Section~\ref{sec:future} we discuss possible research in this
direction.

\section{Preliminaries}
\label{sec:prelim}
In this paper, we consider directed and weighted graphs $G = (V, E, w)$, where $V$
is a set of \emph{vertices}, $E \subseteq V \times V$ a set of \emph{edges}
connecting vertices and $w: E \rightarrow \mathbb{R}_0^+$ a non-negative edge
weight function.
A path is a sequence of vertices $(v_0, \dots , v_n)$ such that $(v_i, v_{i+1})
\in E$ for $0 \leq i < n$. The \emph{length} of a path is the sum of its edge
weights.
The length of a shortest path with
source vertex $s$ and target vertex $t$ is also called the \emph{distance}
between $s$ and $t$, or~$\mydist(s,t)$. 

The classical algorithm for finding shortest paths is Dijkstra's
algorithm~\cite{dijkstra1959note}. It maintains a \emph{distance label} ($\mydist$) for each
vertex and repeatedly \emph{settles} the vertex $u$ with
the currently smallest distance label among all unsettled vertices. It then
\emph{relaxes} all outgoing edges $(u, v)$ by setting $\mydist(v) \gets
\min{(\mydist(v), \mydist(u) + w(u, v))}$.
In the \emph{bidirectional} version of Dijkstra's algorithm, the \emph{forward} search from $s$ is complemented by a \emph{backward} search from $t$ that only considers incoming edges of the settled vertices.

A \emph{shortcut} is an edge whose length corresponds to the length of some nontrivial path in the graph.
For example, for edges $e_1 = (u,v)$ and $e_2 = (v, w)$, a shortcut $e_s = (u, w)$ with
$w(e_s) = w(e_1) + w(e_2)$ can be added to the graph. Note that adding shortcuts
does not change the distance for any pair of vertices in the graph. Also, by
storing skipped vertices, we can recursively \emph{unpack}
shortcuts, e.g., by replacing $e_s$ with $e_1$ and $e_2$ to find the
corresponding path that only uses original (non-shortcut) edges. 

Contraction Hierarchies \cite{GSSD08,geisberger2012exact,dibbelt2016customizable} use
shortcuts to build a hierarchy where every vertex is on its own level. Vertices
are repeatedly removed from the graph in order of a measure of importance. If for any pair of incoming and outgoing
neighbors $u, w$ the removed vertex $v$ is on the \emph{only} shortest path
$(u, v, w)$, then a shortcut $(u, w)$ is added. Whether this shortcut is
necessary is determined by a so-called \emph{witness search} that runs a
shortest path search starting at $u$ on the \emph{overlay graph}. The
overlay graph consists of all vertices not yet removed and all edges incident to
these vertices. The witness search can be restricted to stop after settling a
small amount of vertices. This might add unnecessary shortcuts but does not affect
correctness, while having the potential to speed up the algorithm. Vertex
importance is usually determined by a combination of different measures.
Metrics successfully implemented in previous work (and used in the implementation we
compare against in our evaluation) are the amount of shortcuts added when a vertex were removed
next, the number of \emph{hops} represented by these shortcuts and an additional
\emph{level} metric that helps removing vertices uniformly throughout the
graph. These numbers have in common that they only change when a neighbor of a
vertex is removed from the graph. The algorithm therefore maintains all vertices
in a priority queue with their importance as key. When a vertex is removed, the
importance of its neighbors are updated.
The query
algorithm is a bidirectional Dijkstra search that only relaxes edges that
connect a vertex to a more (less) important vertex in the forward (backward) search. Due to this, edges only need to be
stored at the end point that is removed first.

\section{More Related Work}
\label{sec:relatedwork}

There has been a lot of work on route planning.  Refer to
\cite{bast2016route} for a recent overview.  Here we only give
selected references to place EHs into the big picture.  Besides
hierarchical route planning techniques there are also
techniques which direct the shortest path search towards the
goal (e.g., landmarks \cite{GKW06}, precomputed cluster
distances \cite{MSM07}, arc flags \cite{MSSWW05}). On road networks
goal directed techniques are usually inferior to hierarchical ones since
they need considerably more query or preprocessing time.
However, combining goal directed and hierarchical route planning is a
useful approach \cite{GKW06,BauerDSSSW10}.  We expect that this is
also possible for EHs using the same techniques as used before.  Other
techniques allow very fast queries by building shortest paths directly from two
(hub labeling \cite{AbrahamDGW11}) or three (transit node routing
\cite{BFSS07,ArzLS13}) precomputed shortcuts without requiring a graph
search.  However, these methods require considerably more space than
EHs.

\section{Edge Hierarchies}
\label{sec:main}

The main idea of EHs is to provide a precomputed data structure
that allows queries similar to those of CHs: All shortest
paths can be found by a bidirectional Dijkstra search that only searches ``upwards''. In
contrast to CHs, which build a hierarchy of vertices, EHs build a hierarchy of edges.
Let $r(u,v)$ denote the \emph{rank} assigned to
the edge $(u,v)$. Then, paths found by an EH
query have the form $(s= v_0,\dots , v_m, \dots , v_n = t)$
with $r(v_{i - 1}, v_i) \leq r(v_{i}, v_{i + 1})$
for $0 < i \leq m$ and $r(v_{i - 1}, v_i) \geq r(v_{i}, v_{i + 1})$ for $m < i < n$ 
(allowing $s = m$ or $ t = m$). In line with the terminology from CHs, we call such paths \emph{up-down paths}.

\begin{figure}[t]
  \centering
  \includegraphics[page=3]{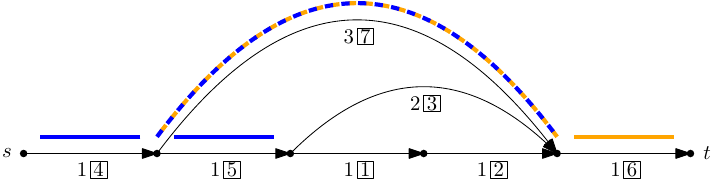}
  \caption{Search space of an EH Query. Blue edges are in the search space of
    the forward search, orange edges are in the search space of the backward
    search. Boxed numbers are edge ranks, unboxed numbers are edge weights.}
  \label{fig:searchSpace}
\end{figure}

The EH query is a modified version of the bidirectional variant of Dijkstra's
algorithm:
In addition to the distance label $\mydist$, we maintain a rank label $r$ at
every node, set to $0$ for $s$ and $t$. When settling a vertex $u$, only edges
with $r(u, v) \geq r(u)$ are relaxed. Whenever $\mydist(v)$ is updated while
relaxing an edge $(u, v)$, $r(v)$ is set to $r(u, v)$. For a stopping condition,
the algorithm maintains an upper bound $\dupper$
for $\mydist(s,t)$ (initially $\infty$) which is updated whenever
a vertex is settled that has already been settled from the other direction.
No edges leaving vertices with $\mydist(v)>\dupper$ are relaxed.
Figure~\ref{fig:searchSpace} illustrates the search space of an Edge Hierarchy
Query. Note how the edges ranked $2$ and $3$ are not in the search space of the backward
search, even though their target vertex is settled.

Algorithm~\ref{algo:construction} shows an algorithm template for constructing an
EH. Initially, all edges are unranked (which we will treat as rank
$\infty$). In iteration $i$, we pick an unranked edge $(u, v)$ and set
its rank to $i$ . We then
iterate over all unranked edges $(u',u)$ and $(v,v')$ and test whether $(u', u, v, v')$ is a shortest path. If yes, we add
either $(u', v)$ or $(u, v')$ as a shortcut. If either of these two edges
already exists, we instead adjust its weight and reset its rank to $\infty$, if it
was already ranked before.

\begin{algorithm}[t]
  currentRank $\gets 0$\;
  \While{\textnormal{Unranked edges remain}} {
    Pick unranked edge $(u, v)$\;
    $r(u, v) \gets \text{currentRank}\Increment$\;
    \For{\textnormal{all unranked edges $(u', u)$}} {
      \For{\textnormal{all unranked edges $(v, v')$}} {
        \If{$\mydist(u', v')$ $= w(u', u) + w(u, v) + w(v, v')$} {
          Add shortcut $(u', v)$ or $(u, v')$\tcp*{Or adjust weight + unset rank}
        }
      }
    }
  }
  \caption{BuildEdgeHierarchy}
  \label{algo:construction}
\end{algorithm}

\begin{theorem}
For every pair of vertices $s$ and $t$, such that there is a path from $s$ to
$t$ in the input graph, Algorithm~\ref{algo:construction}
assigns ranks and adds shortcuts such that there is a shortest up-down path from
$s$ to $t$.
\end{theorem}

\begin{proof}
  We prove this by showing the following: If at the beginning of iteration $i$,
  there is a shortest path from $s$ to $t$ that only uses unranked edges, then
  in iteration $j > i$, there exists an up-down-path $p$ from $s$ to $t$ that only
  uses edges of rank $\geq i$. As at the beginning of the first
  iteration, all edges are unranked, this proves the theorem.

  In iteration $i$, an edge $e$ gets ranked. Let $p$ be a shortest path from $s$
  to $t$ consisting only of unranked edges. If $e$ is not part of $p$,
  then $p$ is still a shortest path that only uses unranked (rank $\infty$)
  edges (which is an up-down path by definition).

  If $e$ is at neither end of $p$, then a shortcut is inserted that replaces two
  edges of $p$, so there still is a shortest path only using unranked edges from $s$ to
  $t$. 

  If $e = (s, v)$ (the case $e = (v, t)$ is analogous) we distinguish two cases:
  \begin{enumerate}
    \item There still exists a shortest path of unranked edges from $s$ to $v$: Then
      there is also a shortest path of unranked edges from $s$ to $t$.
    \item There is no shortest path of unranked edges from $s$ to $v$: Then $(s, v)$ gets
      assigned rank $i$ and can
      never change its rank (note for this, that edges can only be inserted or
      assigned to a different rank if
      there is a shortest path of unranked edges between their endpoints). Furthermore, there is a
      shortest path of
      unranked edges from $v$ to $t$. By induction, in every iteration $j > i$, there will be an up-down-path from $v$ to $t$
      that uses only edges of rank $\geq i$. By adding the edge $(s,
      v)$ to the beginning of that path, we get an up-down path from $s$ to $t$.
  \end{enumerate}
  As the induction basis, note that at the end of the algorithm, no edges
  are unranked, so the claim holds trivially.
\end{proof}

Note that from the induction in the proof above, it follows that we can use the
EH query for the distance calculation in Algorithm~\ref{algo:construction}.

The algorithm can also be slightly altered by only adding a shortcut if $(u', u,
v, v')$ is the \emph{only} remaining unranked shortest path from $u'$ to $v'$. However,
preliminary experiments showed that the version presented here yields better results.

\begin{figure}
  \centering
  \includegraphics{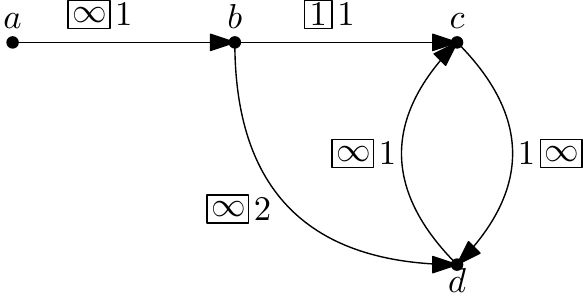}
  \caption{Example showing that EH construction needs to calculate distances on
    the complete graph. Boxed numbers are edge ranks, unboxed numbers are edge weights.}
  \label{fig:needQueryGraph}
\end{figure}

An important difference to CH construction is that
Algorithm~\ref{algo:construction} has to calculate distances in the complete
graph, whereas CH construction only has to query the overlay graph. See Figure~\ref{fig:needQueryGraph} for an
example why using the overlay graph does not suffice for EHs: If $(b, d)$ is assigned
rank $2$, we need to check whether $p = (a,b,d,c)$ is a shortest path. If only
the overlay graph were used for the distance calculation, then we would falsely
assume that $p$ is a shortest path and add a shortcut.

\subsection{Shortcut Selection}
\begin{figure}
  \centering
  \includegraphics{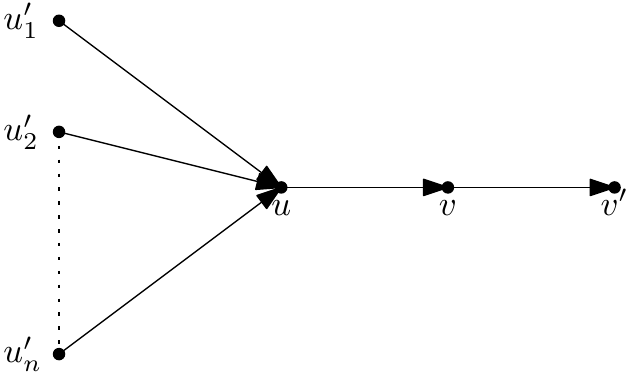}
  \caption{When ranking $(u, v)$, we could either add \emph{all} shortcuts
    $(u'_i, v)$ or just $(u, v')$.}
  \label{fig:lessShortcuts}
\end{figure}
The choice of
the shortcut that is added in the inner loop of
Algorithm~\ref{algo:construction} can make a significant impact on the total
number of shortcuts added. For example, in Figure~\ref{fig:lessShortcuts}, we could either add the
shortcut $(u, v')$ or \emph{all} of the shortcuts $(u'_i,v)$ (assuming $(u'_i,
u, v, v')$ is a shortest path for all $u'_i$). In contrast,
in CHs there is no choice of which shortcut to add.
We minimize the
number of shortcuts added using a solution to a minimum
bipartite vertex cover problem for every iteration of the outer while-loop of Algorithm~\ref{algo:construction}.

The problem $(U \cup V, E)$ is constructed as followed: Instead of directly adding one of the
two possible shortcuts, we add the vertices $u', v'$ to $U, V$ respectively (if
they have not been added before) and
an edge between them.

After all shortcut candidates for an iteration of the
outer loop have been added to the bipartite graph, we compute a minimum Vertex
Cover $C$. Note that this can be done in polynomial time via maximum cardinality
bipartite matching using König's Theorem. We then add the shortcuts $(u', v)$ for
every $u' \in U \cap C$ and $(u, v')$ for every $v' \in V \cap C$. It is easy to
verify that for every pair of candidate shortcuts, one is added. Also, every set
of shortcuts added implies a Vertex Cover for the
graph above, so finding a \emph{minimum} Vertex Cover minimizes the number of
shortcuts added in every iteration of the construction algorithm, given the edge
that is assigned a rank.

To further minimize the number of shortcuts added, we always prefer edges already present in the graph: if $(u',
v)$ or $(u, v')$ is already in the graph (ranked or unranked), we change its
weight accordingly and reset its rank. The pair $(u', v')$ is then not added to
the minimum Vertex Cover problem described above.
\subsection{Edge Selection}

In every iteration of Algorithm~\ref{algo:construction}, an edge is selected to
rank. Our heuristic to select these edges is guided by two goals: Adding a small
number of shortcut edges to the graph, and ranking edges uniformly throughout
the graph. Here, we present the version that produced the best results in our
preliminary experiments. Other versions that resemble the vertex
selection strategies used for CHs resulted in worse
preprocessing and query times.

Our heuristic works in rounds: in  the beginning of each round, a set of edges to rank is selected
and fixed. Only when all edges selected are ranked, a new
round is started and a new set of edges is selected. Edges are selected by
counting for each unranked edge $e$
the number of new shortcuts that would be added if $e$ was ranked.
This is done by simulating an iteration of the outer while-loop of
Algorithm~\ref{algo:construction} without actually adding any shortcuts to the graph and resetting $r(u,v)$ to $\infty$ afterwards. Then, we select all edges that cause the minimum
number of shortcuts among all their incident edges.

\subsection{Stalling}
\label{sec:stalling}

A technique that significantly reduces query times for CHs
is called \emph{Stall on Demand}. The idea is to stall the search at vertices
that do not lie on a shortest path from $s$ to $t$ by checking whether a shorter
path can be found via incoming (outgoing) downward edges in the forward
(backward) search. This can happen because CHs only
guarantee shortest up-down paths between any pairs of vertices. The same is true
for EHs. We present two stalling techniques that can be used
with EHs.

\begin{description}
\item[Stall on Demand]
In EHs any edge can be a downward or an upward edge depending on
the rank of the edges leading to the source vertex of that edge. Stall on Demand
checks \emph{all} incoming (outgoing) edges in the forward (backward) search.

\item[Stall in Advance]
Stall on Demand may relax every edge twice: Once
when settling the source (target) vertex and once for stalling when settling the
target (source)
vertex in the forward (backward) search. \emph{Stall in Advance} relaxes
every edge at most once: when settling a vertex $u$, we not only relax all
outgoing (incoming)
edges that are ranked higher than the path to $u$, but also all edges that are
ranked lower. However, we do not update $\mydist$ with the distance computed via the low ranked
edges. Instead, we store it in a separate $\mystalldist$
label. To check whether we can stall the search at vertex $v$, we compare $\mydist(v)$ with $\mystalldist(v)$. If
$\mystalldist$ is smaller, we can
stall at $v$.
\end{description}

\section{Experimental Evaluation}\label{sec:experiments}

We implement EHs in C++ and compile with gcc 7.4.0 using full
optimizations~(\texttt{-O3}). Our implementation of the construction algorithm
is relatively straight forward without much emphasis on optimizations. For
queries, we use adjacency arrays for incoming and outgoing edges and sort all
edges incident to a vertex in descending order of their rank. This way we can
stop iterating over a vertex's neighborhood once we find an edge with a lower rank
than allowed for the current path. Additionally, we reorder the vertices in
depth-first-search-preorder for better memory locality.
The EH \emph{construction} algorithm uses CH queries
to find the distance between two vertices. 
The source code is available on GitHub\footnote{\url{https://github.com/Hespian/EdgeHierarchies}}.

For comparison with CHs, we
use the implementation from
RoutingKit\footnote{\url{https://github.com/RoutingKit/RoutingKit}}~\cite{dibbelt2016customizable}
where queries use Stall on Demand. 

The machine used for all experiments is equipped with 4 x Intel Xeon
E5-4640 at 2.4 GHz and 512 GiB DDR3-PC1600 RAM but only a single core is utilized.

\subsection{Data Sets}
We evaluate EHs on two benchmark graphs from the DIMACS Challenge on
Shortest Paths~\cite{demetrescu2009dimacs}: The road network of Western Europe from PTV AG
with 18 million vertices and 42 million directed edges, and the TIGER/USA
road network with 23 million vertices and 29 million undirected edges
(resulting in 58 million directed edges), as well as smaller subsets of the
TIGER/USA graph. Both graphs are available with edge weights corresponding to
travel times or geographic distance.

\begin{figure}
  \centering
  \includegraphics{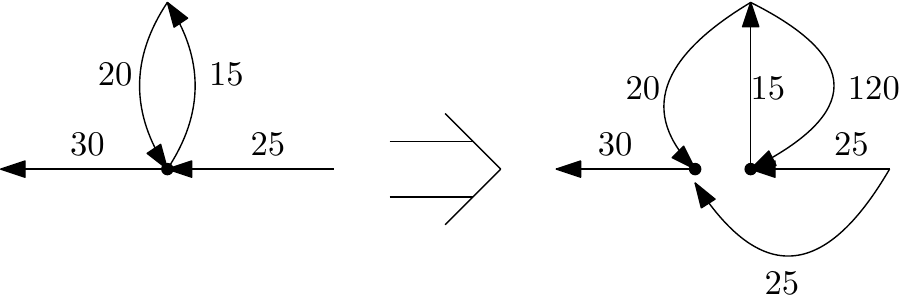}
  \caption{Left: Original graph. Right: Graph with added turns. 100 seconds are
    added to the edge corresponding to a U-turn.}
  \label{fig:turnCosts}
\end{figure}

In addition to these graphs, we also evaluate the performance on graphs that
model the cost for taking turns at a crossing.
We follow the approach used in \cite{delling2015customizable, bast2016route}
to define simple turn costs that reportedly yield performance characteristics
similar to truly realistic
values:
For the travel time metric, we assign
costs of 100 seconds for U-turns (meaning an edge pair $(u,v), (v,u)$) and 0 for
all other turns. For the distance metric, all turns are free.
We explicitly
model turns into our graphs. This can be done by splitting every vertex $v$
into a number of vertices equal to its degree and connecting each new vertex to one of
$v$'s incident edges. Then, edges between the new vertices are added: For each
vertex incident to one of $v$'s incoming edges, an edge is added to each of the
vertices incident to one of $v$'s outgoing edges. The weights of these new edges
are set to the turn costs. We use a more compact representation of the same
concept: We only split a vertex into a number of vertices equal to its outgoing
degree and connect incoming edges directly to these new vertices, adding the
turn costs to the edge weights.
Figure~\ref{fig:turnCosts} shows an example for travel times. Table~\ref{tab:instances} lists all instances and their sizes used in our evaluation.

The distance metric as well as adding turn information are cases in which CHs
were shown to perform significantly worse than with the travel time metric and
without turn information (e.g.~\cite{delling2015customizable}).

\begin{table}[h]
  \centering
  \caption{Instances used in our evaluation. \emph{With turns} are original
    instances with added turns.}
  \label{tab:instances}
\begin{tabular}{l|rr|rr}
                             & \multicolumn{2}{c|}{Original}                          & \multicolumn{2}{c}{With turns}                        \\
  \multicolumn{1}{c|}{Graph} & \multicolumn{1}{c}{$|V|$} & \multicolumn{1}{c|}{$|E|$} & \multicolumn{1}{c}{$|V|$} & \multicolumn{1}{c}{$|E|$} \\
  \hline
  USA.BAY                    & \numprint{321270}         & \numprint{794830}          & \numprint{794830}         & \numprint{2279208}        \\
  USA.W                      & \numprint{6262104}        & \numprint{15119284}        & \numprint{15119284}       & \numprint{41815474}       \\
  USA.CTR                    & \numprint{14081816}       & \numprint{33866826}        & \numprint{33866826}       & \numprint{93609832}       \\
  USA                        & \numprint{23947347}       & \numprint{57708624}        & \numprint{57708624}       & \numprint{159734066}      \\
  EUROPE                     & \numprint{18010173}       & \numprint{42188664}        & \numprint{42188664}       & \numprint{113953602}      \\
\end{tabular}
\end{table}

\subsection{Choosing the Right Stalling Technique}
\label{sec:experiments:stalling}

In this section we evaluate the stalling techniques explained in
Section~\ref{sec:stalling}. To get some insight in how stalling performs for
other techniques, we compare to Stall on Demand for CHs.
Tables~\ref{tab:StallingTime} and \ref{tab:StallingDistance} compare the query times, number of vertices settled and edges
relaxed for different stalling techniques averaged over \numprint{100000} random queries. The number of edges actually relaxed and
the number of edges ``relaxed'' to check whether the search can be stalled are
shown separately. We also count the number of vertices that are settled at their
actual distance to the source vertex (min. vertices). This gives an insight into how many vertices
would be settled with a \emph{perfect} stalling technique. For the travel time metric,
EHs with both Stall on Demand and Stall in Advance perform more
stall checks than CHs, outweighing the savings in number of
vertices settled and leading to longer query times than without any stalling. For the distance metric, Stall on Demand reduces the number
of vertices settled for EHs to less than for CHs. The total of number of edges
touched is also less for EHs. However, running times are still faster without
stalling because less edges are relaxed (or considered for stalling) and thus less distance labels are touched. Due to the additional distance label, Stall in Advance significantly
increases query times. The last column also shows that stalling holds
more potential for CHs than for EHs. However, we also see that EHs already
perform relatively well without stalling: CHs on the travel time metric would have to settle more than
twice as many vertices as EHs if no stalling was used and even when not counting the
stall checks, CHs with Stall on Demand relax more edges than EHs. For the
distance metric, this is even more severe: Here, the search space for CHs
without Stall on Demand increases so much that query times increase to over 3
ms. EHs already settle a reasonably small number of vertices without stalling.

\begin{figure}[t]
  \centering
  \input{images/partialStalling.pgf}
  \caption{Speedup of query with partial stalling over unstalled query with different fractions of
    edges used for stalling. Times were measured on the EUROPE road network.}
\label{fig:partialStalling}
\end{figure}
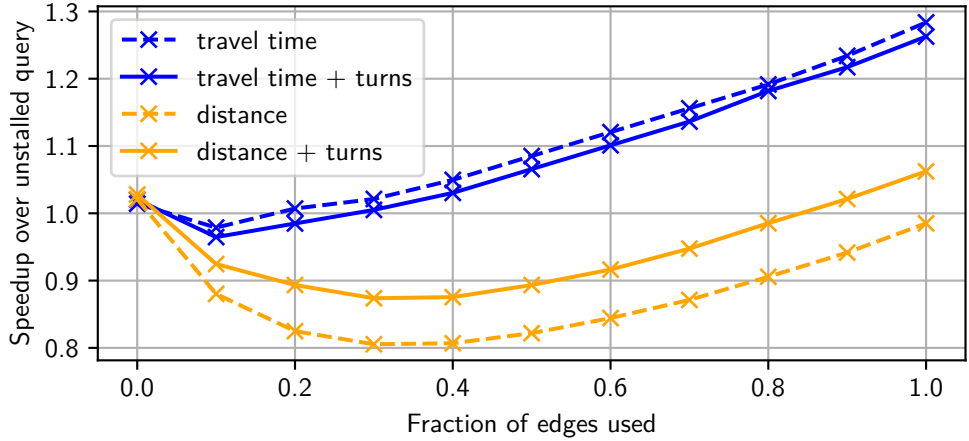

These experiments show that the increased number of edges touched
outweighs the decreased number of vertices settled. Thus, a stalling technique that
only touches \emph{some} more edges might lead to improved running times if it
successfully stalls at enough vertices.
Figure~\ref{fig:partialStalling} shows the performance when only
a fraction of the edges incident to a vertex are considered for Stall on Demand -- going from high
ranked edges to low ranked edges (note that this can be done efficiently in our
implementation as edges are stored ordered by their rank). We are going to refer
to this as \emph{partial stalling} from here on. We see a slight increase in
running time due to the associated calculations (see the data point for 
$x = 0.0$) but all instances shown benefit from partial stalling for some
fraction ($10\%$ for travel times and $30\%$ for distances).

\begin{table}[t]
\centering
\caption{Query results for different stalling techniques for Edge
  Hierarchies and Contraction Hierarchies on the EUROPE road
  network with the \textbf{travel time} metric and \textbf{turns}.}
\label{tab:StallingTime}
\begin{tabular}{c|c|r|r|r|r|r}
  Algo.                                                          & Stalling      & time [$\mu$s] & settled & relaxed & stall checks & min. vertices          \\
  \hline
  \parbox[t]{2mm}{\multirow{3}{*}{\rotatebox[origin=c]{90}{EH}}} & -             & 199           & 906     & 1734    & -            & \multirow{3}{*}{ 361 } \\
                                                                 & S. on Demand  & 250           & 604     & 958     & 11920        &                        \\
                                                                 & S. in Advance & 471           & 614     & 982     & 10563        &                        \\
  \hline
  \parbox[t]{2mm}{\multirow{2}{*}{\rotatebox[origin=c]{90}{CH}}} & S. on Demand  & 130           & 533     & 1969    & 2888         & \multirow{2}{*}{ 253 } \\
                                                                 & -             & 338           & 1996    & 15500   & -            &                        \\
\end{tabular}
\end{table}

\begin{table}[t]
  \centering
  \caption{Query results for different stalling techniques for Edge
    Hierarchies and Contraction Hierarchies on the EUROPE road
    network with the \textbf{distance} metric and \textbf{turns}.}
  \label{tab:StallingDistance}
  \begin{tabular}{c|c|r|r|r|r|r}
    Algo.                                                          & Stalling      & time [$\mu$s] & settled & relaxed & stall checks & min. vertices          \\
    \hline
    \parbox[t]{2mm}{\multirow{3}{*}{\rotatebox[origin=c]{90}{EH}}} & -             & 608           & 2573    & 5586    & -            & \multirow{3}{*}{ 638 } \\
                                                                   & S. on Demand  & 642           & 1368    & 2276    & 29192        &                        \\
                                                                   & S. in Advance & 1387          & 1439    & 2442    & 26959        &                        \\
    \hline
    \parbox[t]{2mm}{\multirow{2}{*}{\rotatebox[origin=c]{90}{CH}}} & S. on Demand  & 634           & 1943    & 16849   & 25007        & \multirow{2}{*}{ 704 } \\
                                                                   & -             & 3403          & 12320   & 300758  & -            &                        \\
  \end{tabular}

\end{table}

\subsection{Main Results}
\label{sec:results}

As EHs share similarities with CHs,
both using similar query algorithms, we compare the two with respect to their
preprocessing and query times as well as the number of vertices settled and
edges relaxed during queries. Another interesting property is the number of
edges in the hierarchy. Note however, that CHs only store each edge
once, whereas EHs need to store each edge at both endpoints.
Tables~\ref{tab:OverviewTravelTime} and~\ref{tab:OverviewDistance} show these numbers
averaged over \numprint{100000} random queries. We execute queries without Stall
on Demand and with partial stalling in increments of $10\%$. The numbers reported
here are for the best query times among these stalling configurations as
indicated by the last column. In a
real-world system the optimal configuration could be found as a part of the
preprocessing step. Due to time restrictions, the construction was
only run once for each algorithm and instance. Checking whether the search can be stalled at
a vertex is essentially an edge relaxation (minus priority queue
operations), so we combine these numbers here. We can see that EHs suffer less
from adding turns to the graphs than CHs. While the number of
shortcuts added is comparable for EHs and CHs on the original graphs (with CHs even adding slightly
fewer), CHs add significantly more when turns are added. This can also be seen in
the number of edges relaxed: The number of edges relaxed with and without turns
are very similar for EHs. For the distance metric, EHs perform even better
when adding turns than on the original inputs. With turns, EHs almost always
relax less than half as many edges as CHs. This shows that the intuition
behind EHs -- ranking \emph{roads} (edges) rather than \emph{junctions}
(vertices) -- helps to better prune roads that are irrelevant for the
query. However, CHs usually settle between 2 and 3 times less vertices (except
for the distance metric with turns where EHs often settle less vertices than CHs). Overall
this leads to longer query times for EHs in most cases. For the distance metric with turns, query times for EHs are close to CHs
-- for the EUROPE instance EHs even achieve faster queries.
The preprocessing step is much faster for CHs, partially
due to our unoptimized implementation, but the CH vertex ranking also only
updates the neighbors of a vertex after it was ranked. The edge ranking we use, on
the other hand, simulates the ranking of every edge for each round of edge
selection. The CH implementation in RoutingKit also limits the number of steps
done for the witness search, giving additional speed up. As EHs have to
find witnesses and (depending on the edge ranking technique) calculate
importance values
for every edge, compared to CHs having
to do the same for every vertex, longer preprocessing times are to be expected.

\begin{table}[t]
\centering
\caption{Running times and search space sizes of Edge Hierarchies and
  Contraction Hierarchies on different graphs with the \textbf{travel time} metric.}
\label{tab:OverviewTravelTime}
\addtolength{\tabcolsep}{-1pt}    
\begin{tabular}{cc|rr|rr|rr|rr|rr|c}
  \multicolumn{2}{c|}{\multirow{2}{*}{Graph}}                            & \multicolumn{2}{c|}{Prepr. [s]} & \multicolumn{2}{c|}{|E| [M]} & \multicolumn{2}{c|}{Query [$\mu$s]} & \multicolumn{2}{c|}{settled} & \multicolumn{2}{c|}{relaxed} &  \multicolumn{1}{c}{stall.} \\
  \multicolumn{2}{c|}{}                                                            & EH    & CH    & EH    & CH    & EH  & CH  & EH   & CH  & EH   & CH & $\%$         \\
  \hline
  \parbox[t]{2mm}{\multirow{5}{*}{\rotatebox[origin=c]{90}{Original}}}   & USA.BAY & 100   & 6     & 1.4   & 1.4   & 37  & 16  & 301  & 108 & 710  & 679  & - \\ 
                                                                         & USA.W   & 1785  & 153   & 27.5  & 27.4  & 96  & 37  & 538  & 193 & 1299 & 1386 & - \\ 
                                                                         & USA.CTR & 4389  & 482   & 61.5  & 61.1  & 140 & 53  & 612  & 254 & 3132 & 2136 & 10 \\ 
                                                                         & USA     & 7145  & 674   & 104.5 & 104.0 & 153 & 60  & 643  & 271 & 3320 & 2253 & 10 \\ 
                                                                         & EUROPE  & 3171  & 453   & 70.3  & 70.3  & 138 & 75  & 607  & 356 & 2443 & 2967 & 10 \\ 
  \hline
  \parbox[t]{2mm}{\multirow{5}{*}{\rotatebox[origin=c]{90}{With turns}}} & USA.BAY & 634   & 156   & 4.0   & 6.0   & 79  & 67  & 511  & 362 & 929  & 3253 & - \\ 
                                                                         & USA.W   & 9403  & 2730  & 69.9  & 105.1 & 165 & 124 & 748  & 564 & 1365 & 4810 & - \\ 
                                                                         & USA.CTR & 25084 & 7316  & 159.3 & 239.2 & 240 & 172 & 885  & 700 & 3126 & 6530 & 10 \\ 
                                                                         & USA     & 45904 & 15462 & 270.3 & 404.3 & 250 & 186 & 900  & 737 & 3217 & 6792 & 10 \\ 
                                                                         & EUROPE  & 17822 & 4743  & 194.0 & 249.1 & 191 & 130 & 726  & 533 & 2662 & 4857 & 10 \\ 
\end{tabular}
\addtolength{\tabcolsep}{1pt}    
\end{table}

\begin{table}[t]
\centering
\caption{Running times and search space sizes of Edge Hierarchies and
  Contraction Hierarchies on different graphs with the \textbf{distance} metric.}
\label{tab:OverviewDistance}
\addtolength{\tabcolsep}{-2pt}    
\begin{tabular}{cc|rr|rr|rr|rr|rr|c}
  \multicolumn{2}{c|}{\multirow{2}{*}{Graph}}                            & \multicolumn{2}{c|}{Prepr. [s]} & \multicolumn{2}{c|}{|E| [M]} & \multicolumn{2}{c|}{Query [$\mu$s]} & \multicolumn{2}{c|}{settled} & \multicolumn{2}{c|}{relaxed}   &  \multicolumn{1}{c}{stall.}                  \\
  \multicolumn{2}{c|}{}                                                            & EH    & CH    & EH    & CH    & EH  & CH  & EH   & CH   & EH   & CH  & $\%$         \\
  \hline
  \parbox[t]{2mm}{\multirow{5}{*}{\rotatebox[origin=c]{90}{Original}}}   & USA.BAY & 166   & 9     & 1.5   & 1.5   & 73  & 30  & 560  & 180  & 1440 & 1686 & -  \\ 
                                                                         & USA.W   & 3435  & 243   & 28.6  & 28.5  & 254 & 96  & 1002 & 446  & 8183 & 6045 & 20 \\ 
                                                                         & USA.CTR & 13062 & 1157  & 65.7  & 65.5  & 526 & 216 & 1697 & 832  & 20041 & 15561 & 30 \\ 
                                                                         & USA     & 21041 & 1537  & 110.8 & 110.7 & 573 & 235 & 1769 & 897  & 21461 & 16787 & 30 \\ 
                                                                         & EUROPE  & 14487 & 2152  & 79.6  & 79.6  & 538 & 355 & 1756 & 1179 & 19793 & 27807 & 30 \\ 
  \hline
  \parbox[t]{2mm}{\multirow{5}{*}{\rotatebox[origin=c]{90}{With turns}}} & USA.BAY & 476   & 158   & 3.6   & 5.7   & 95  & 92  & 623  & 470  & 1149 & 4979  & - \\ 
                                                                         & USA.W   & 8452  & 3338  & 64.9  & 102.3 & 278 & 250 & 1289 & 993  & 2564 & 13402 & - \\ 
                                                                         & USA.CTR & 30313 & 13629 & 148.5 & 235.7 & 556 & 537 & 1477 & 1743 & 15286 & 31629 & 40 \\ 
                                                                         & USA     & 58025 & 30869 & 251.1 & 398.3 & 604 & 580 & 1605 & 1849 & 13712 & 33436 & 30 \\ 
                                                                         & EUROPE  & 24757 & 13266 & 172.3 & 267.2 & 533 & 634 & 1543 & 1943 & 13355 & 41856 & 30 \\ 
\end{tabular}
\addtolength{\tabcolsep}{2pt}    
\end{table}

The random queries used for the experiments above are long-ranged on average.
However, real-world queries tend to be short-ranged. For this reason, Sanders
and Schultes~\cite{sanders2005highway} introduce an evaluation methodology using
\emph{Dijkstra Ranks}. When running a Dijkstra query starting at some vertex in
the graph, the $i$th vertex removed from the priority queue is assigned Dijkstra
Rank $i$. Figures~\ref{fig:dijkstraRankTime} and \ref{fig:dijkstraRankDistance}
show the number of vertices
settled, number of edges relaxed, and query times for vertices of Dijkstra Ranks
$2^6, \dots ,
2^{\lfloor \log |V| \rfloor}$ from \numprint{1000} random starting vertices. This way, the performance of algorithms can be
observed for both short-ranged and long-ranged queries (and everything in
between). EHs use $10\%$ and  $30\%$ partial stalling for travel times
and distances, respectively. The comparison between number of vertices settled
and query time shows that the algorithm that settles less vertices has the
faster query time and edge relaxations play a less important role. This is
likely due to vertex accesses causing more cache misses than accesses to the
edges of a singe vertex. If one would improve the cache
efficiency by better node orderings or other improvements, it seems possible that EHs
decreased number of relaxed edges can outweigh the increased number of settled vertices.

\begin{figure}[t!]
    \centering
    \input{images/scc-eur2time.grTurnCostsDijkstraDFSPreOrderEHBackwardStallingPartialStalling10.pgf}
  \caption{Number of vertices settled and edges relaxed, and query times for different Dijkstra
    Ranks on EUROPE with the \textbf{travel time} metric and \textbf{turns}.}\label{fig:dijkstraRankTime}

  \input{images/scc-eur2dist.grTurnCostsDijkstraDFSPreOrderEHBackwardStallingPartialStalling30.pgf}
  \caption{Number of vertices settled and edges relaxed, and query times for different Dijkstra
    Ranks on EUROPE with the \textbf{distance} metric and \textbf{turns}.}\label{fig:dijkstraRankDistance}
\end{figure}

\section{Future Work}\label{sec:future}
For CHs there is a lot of experience with configuring the
preprocessing process. The additional complications of EH
preprocessing make it likely that much better versions are possible also for EHs.
Trying different ways of cleaning up the distance labels for
new queries might lead to some improvements as preliminary experiments
showed some effect here. Due to EHs being less cache-efficient than CHs right
now, we expect them to profit more from such changes. On the application side, we can look for networks
with different characteristics where EHs might have advantages. For
road networks, we might harvest the advantage in number of relaxed
edges by looking at generalizations of static shortest path search
where edge relaxations are expensive, e.g., time-dependent edge costs
\cite{BatzGSV13,KontogiannisWZ16} or multicriteria shortest paths.

\clearpage


\bibliography{references}

\end{document}

%% file: images/partialStalling.pgf
\begingroup%
\makeatletter%
\begin{pgfpicture}%
\pgfpathrectangle{\pgfpointorigin}{\pgfqpoint{5.188194in}{2.485638in}}%
\pgfusepath{use as bounding box, clip}%
\begin{pgfscope}%
\pgfsetbuttcap%
\pgfsetmiterjoin%
\definecolor{currentfill}{rgb}{1.000000,1.000000,1.000000}%
\pgfsetfillcolor{currentfill}%
\pgfsetlinewidth{0.000000pt}%
\definecolor{currentstroke}{rgb}{1.000000,1.000000,1.000000}%
\pgfsetstrokecolor{currentstroke}%
\pgfsetdash{}{0pt}%
\pgfpathmoveto{\pgfqpoint{0.000000in}{0.000000in}}%
\pgfpathlineto{\pgfqpoint{5.188194in}{0.000000in}}%
\pgfpathlineto{\pgfqpoint{5.188194in}{2.485638in}}%
\pgfpathlineto{\pgfqpoint{0.000000in}{2.485638in}}%
\pgfpathclose%
\pgfusepath{fill}%
\end{pgfscope}%
\begin{pgfscope}%
\pgfsetbuttcap%
\pgfsetmiterjoin%
\definecolor{currentfill}{rgb}{1.000000,1.000000,1.000000}%
\pgfsetfillcolor{currentfill}%
\pgfsetlinewidth{0.000000pt}%
\definecolor{currentstroke}{rgb}{0.000000,0.000000,0.000000}%
\pgfsetstrokecolor{currentstroke}%
\pgfsetstrokeopacity{0.000000}%
\pgfsetdash{}{0pt}%
\pgfpathmoveto{\pgfqpoint{0.558194in}{0.502638in}}%
\pgfpathlineto{\pgfqpoint{5.053194in}{0.502638in}}%
\pgfpathlineto{\pgfqpoint{5.053194in}{2.350638in}}%
\pgfpathlineto{\pgfqpoint{0.558194in}{2.350638in}}%
\pgfpathclose%
\pgfusepath{fill}%
\end{pgfscope}%
\begin{pgfscope}%
\pgfpathrectangle{\pgfqpoint{0.558194in}{0.502638in}}{\pgfqpoint{4.495000in}{1.848000in}}%
\pgfusepath{clip}%
\pgfsetrectcap%
\pgfsetroundjoin%
\pgfsetlinewidth{0.803000pt}%
\definecolor{currentstroke}{rgb}{0.690196,0.690196,0.690196}%
\pgfsetstrokecolor{currentstroke}%
\pgfsetdash{}{0pt}%
\pgfpathmoveto{\pgfqpoint{0.762512in}{0.502638in}}%
\pgfpathlineto{\pgfqpoint{0.762512in}{2.350638in}}%
\pgfusepath{stroke}%
\end{pgfscope}%
\begin{pgfscope}%
\pgfsetbuttcap%
\pgfsetroundjoin%
\definecolor{currentfill}{rgb}{0.000000,0.000000,0.000000}%
\pgfsetfillcolor{currentfill}%
\pgfsetlinewidth{0.803000pt}%
\definecolor{currentstroke}{rgb}{0.000000,0.000000,0.000000}%
\pgfsetstrokecolor{currentstroke}%
\pgfsetdash{}{0pt}%
\pgfsys@defobject{currentmarker}{\pgfqpoint{0.000000in}{-0.048611in}}{\pgfqpoint{0.000000in}{0.000000in}}{%
\pgfpathmoveto{\pgfqpoint{0.000000in}{0.000000in}}%
\pgfpathlineto{\pgfqpoint{0.000000in}{-0.048611in}}%
\pgfusepath{stroke,fill}%
}%
\begin{pgfscope}%
\pgfsys@transformshift{0.762512in}{0.502638in}%
\pgfsys@useobject{currentmarker}{}%
\end{pgfscope}%
\end{pgfscope}%
\begin{pgfscope}%
\definecolor{textcolor}{rgb}{0.000000,0.000000,0.000000}%
\pgfsetstrokecolor{textcolor}%
\pgfsetfillcolor{textcolor}%
\pgftext[x=0.762512in,y=0.405416in,,top]{\color{textcolor}\sffamily\fontsize{10.000000}{12.000000}\selectfont 0.0}%
\end{pgfscope}%
\begin{pgfscope}%
\pgfpathrectangle{\pgfqpoint{0.558194in}{0.502638in}}{\pgfqpoint{4.495000in}{1.848000in}}%
\pgfusepath{clip}%
\pgfsetrectcap%
\pgfsetroundjoin%
\pgfsetlinewidth{0.803000pt}%
\definecolor{currentstroke}{rgb}{0.690196,0.690196,0.690196}%
\pgfsetstrokecolor{currentstroke}%
\pgfsetdash{}{0pt}%
\pgfpathmoveto{\pgfqpoint{1.579785in}{0.502638in}}%
\pgfpathlineto{\pgfqpoint{1.579785in}{2.350638in}}%
\pgfusepath{stroke}%
\end{pgfscope}%
\begin{pgfscope}%
\pgfsetbuttcap%
\pgfsetroundjoin%
\definecolor{currentfill}{rgb}{0.000000,0.000000,0.000000}%
\pgfsetfillcolor{currentfill}%
\pgfsetlinewidth{0.803000pt}%
\definecolor{currentstroke}{rgb}{0.000000,0.000000,0.000000}%
\pgfsetstrokecolor{currentstroke}%
\pgfsetdash{}{0pt}%
\pgfsys@defobject{currentmarker}{\pgfqpoint{0.000000in}{-0.048611in}}{\pgfqpoint{0.000000in}{0.000000in}}{%
\pgfpathmoveto{\pgfqpoint{0.000000in}{0.000000in}}%
\pgfpathlineto{\pgfqpoint{0.000000in}{-0.048611in}}%
\pgfusepath{stroke,fill}%
}%
\begin{pgfscope}%
\pgfsys@transformshift{1.579785in}{0.502638in}%
\pgfsys@useobject{currentmarker}{}%
\end{pgfscope}%
\end{pgfscope}%
\begin{pgfscope}%
\definecolor{textcolor}{rgb}{0.000000,0.000000,0.000000}%
\pgfsetstrokecolor{textcolor}%
\pgfsetfillcolor{textcolor}%
\pgftext[x=1.579785in,y=0.405416in,,top]{\color{textcolor}\sffamily\fontsize{10.000000}{12.000000}\selectfont 0.2}%
\end{pgfscope}%
\begin{pgfscope}%
\pgfpathrectangle{\pgfqpoint{0.558194in}{0.502638in}}{\pgfqpoint{4.495000in}{1.848000in}}%
\pgfusepath{clip}%
\pgfsetrectcap%
\pgfsetroundjoin%
\pgfsetlinewidth{0.803000pt}%
\definecolor{currentstroke}{rgb}{0.690196,0.690196,0.690196}%
\pgfsetstrokecolor{currentstroke}%
\pgfsetdash{}{0pt}%
\pgfpathmoveto{\pgfqpoint{2.397058in}{0.502638in}}%
\pgfpathlineto{\pgfqpoint{2.397058in}{2.350638in}}%
\pgfusepath{stroke}%
\end{pgfscope}%
\begin{pgfscope}%
\pgfsetbuttcap%
\pgfsetroundjoin%
\definecolor{currentfill}{rgb}{0.000000,0.000000,0.000000}%
\pgfsetfillcolor{currentfill}%
\pgfsetlinewidth{0.803000pt}%
\definecolor{currentstroke}{rgb}{0.000000,0.000000,0.000000}%
\pgfsetstrokecolor{currentstroke}%
\pgfsetdash{}{0pt}%
\pgfsys@defobject{currentmarker}{\pgfqpoint{0.000000in}{-0.048611in}}{\pgfqpoint{0.000000in}{0.000000in}}{%
\pgfpathmoveto{\pgfqpoint{0.000000in}{0.000000in}}%
\pgfpathlineto{\pgfqpoint{0.000000in}{-0.048611in}}%
\pgfusepath{stroke,fill}%
}%
\begin{pgfscope}%
\pgfsys@transformshift{2.397058in}{0.502638in}%
\pgfsys@useobject{currentmarker}{}%
\end{pgfscope}%
\end{pgfscope}%
\begin{pgfscope}%
\definecolor{textcolor}{rgb}{0.000000,0.000000,0.000000}%
\pgfsetstrokecolor{textcolor}%
\pgfsetfillcolor{textcolor}%
\pgftext[x=2.397058in,y=0.405416in,,top]{\color{textcolor}\sffamily\fontsize{10.000000}{12.000000}\selectfont 0.4}%
\end{pgfscope}%
\begin{pgfscope}%
\pgfpathrectangle{\pgfqpoint{0.558194in}{0.502638in}}{\pgfqpoint{4.495000in}{1.848000in}}%
\pgfusepath{clip}%
\pgfsetrectcap%
\pgfsetroundjoin%
\pgfsetlinewidth{0.803000pt}%
\definecolor{currentstroke}{rgb}{0.690196,0.690196,0.690196}%
\pgfsetstrokecolor{currentstroke}%
\pgfsetdash{}{0pt}%
\pgfpathmoveto{\pgfqpoint{3.214331in}{0.502638in}}%
\pgfpathlineto{\pgfqpoint{3.214331in}{2.350638in}}%
\pgfusepath{stroke}%
\end{pgfscope}%
\begin{pgfscope}%
\pgfsetbuttcap%
\pgfsetroundjoin%
\definecolor{currentfill}{rgb}{0.000000,0.000000,0.000000}%
\pgfsetfillcolor{currentfill}%
\pgfsetlinewidth{0.803000pt}%
\definecolor{currentstroke}{rgb}{0.000000,0.000000,0.000000}%
\pgfsetstrokecolor{currentstroke}%
\pgfsetdash{}{0pt}%
\pgfsys@defobject{currentmarker}{\pgfqpoint{0.000000in}{-0.048611in}}{\pgfqpoint{0.000000in}{0.000000in}}{%
\pgfpathmoveto{\pgfqpoint{0.000000in}{0.000000in}}%
\pgfpathlineto{\pgfqpoint{0.000000in}{-0.048611in}}%
\pgfusepath{stroke,fill}%
}%
\begin{pgfscope}%
\pgfsys@transformshift{3.214331in}{0.502638in}%
\pgfsys@useobject{currentmarker}{}%
\end{pgfscope}%
\end{pgfscope}%
\begin{pgfscope}%
\definecolor{textcolor}{rgb}{0.000000,0.000000,0.000000}%
\pgfsetstrokecolor{textcolor}%
\pgfsetfillcolor{textcolor}%
\pgftext[x=3.214331in,y=0.405416in,,top]{\color{textcolor}\sffamily\fontsize{10.000000}{12.000000}\selectfont 0.6}%
\end{pgfscope}%
\begin{pgfscope}%
\pgfpathrectangle{\pgfqpoint{0.558194in}{0.502638in}}{\pgfqpoint{4.495000in}{1.848000in}}%
\pgfusepath{clip}%
\pgfsetrectcap%
\pgfsetroundjoin%
\pgfsetlinewidth{0.803000pt}%
\definecolor{currentstroke}{rgb}{0.690196,0.690196,0.690196}%
\pgfsetstrokecolor{currentstroke}%
\pgfsetdash{}{0pt}%
\pgfpathmoveto{\pgfqpoint{4.031603in}{0.502638in}}%
\pgfpathlineto{\pgfqpoint{4.031603in}{2.350638in}}%
\pgfusepath{stroke}%
\end{pgfscope}%
\begin{pgfscope}%
\pgfsetbuttcap%
\pgfsetroundjoin%
\definecolor{currentfill}{rgb}{0.000000,0.000000,0.000000}%
\pgfsetfillcolor{currentfill}%
\pgfsetlinewidth{0.803000pt}%
\definecolor{currentstroke}{rgb}{0.000000,0.000000,0.000000}%
\pgfsetstrokecolor{currentstroke}%
\pgfsetdash{}{0pt}%
\pgfsys@defobject{currentmarker}{\pgfqpoint{0.000000in}{-0.048611in}}{\pgfqpoint{0.000000in}{0.000000in}}{%
\pgfpathmoveto{\pgfqpoint{0.000000in}{0.000000in}}%
\pgfpathlineto{\pgfqpoint{0.000000in}{-0.048611in}}%
\pgfusepath{stroke,fill}%
}%
\begin{pgfscope}%
\pgfsys@transformshift{4.031603in}{0.502638in}%
\pgfsys@useobject{currentmarker}{}%
\end{pgfscope}%
\end{pgfscope}%
\begin{pgfscope}%
\definecolor{textcolor}{rgb}{0.000000,0.000000,0.000000}%
\pgfsetstrokecolor{textcolor}%
\pgfsetfillcolor{textcolor}%
\pgftext[x=4.031603in,y=0.405416in,,top]{\color{textcolor}\sffamily\fontsize{10.000000}{12.000000}\selectfont 0.8}%
\end{pgfscope}%
\begin{pgfscope}%
\pgfpathrectangle{\pgfqpoint{0.558194in}{0.502638in}}{\pgfqpoint{4.495000in}{1.848000in}}%
\pgfusepath{clip}%
\pgfsetrectcap%
\pgfsetroundjoin%
\pgfsetlinewidth{0.803000pt}%
\definecolor{currentstroke}{rgb}{0.690196,0.690196,0.690196}%
\pgfsetstrokecolor{currentstroke}%
\pgfsetdash{}{0pt}%
\pgfpathmoveto{\pgfqpoint{4.848876in}{0.502638in}}%
\pgfpathlineto{\pgfqpoint{4.848876in}{2.350638in}}%
\pgfusepath{stroke}%
\end{pgfscope}%
\begin{pgfscope}%
\pgfsetbuttcap%
\pgfsetroundjoin%
\definecolor{currentfill}{rgb}{0.000000,0.000000,0.000000}%
\pgfsetfillcolor{currentfill}%
\pgfsetlinewidth{0.803000pt}%
\definecolor{currentstroke}{rgb}{0.000000,0.000000,0.000000}%
\pgfsetstrokecolor{currentstroke}%
\pgfsetdash{}{0pt}%
\pgfsys@defobject{currentmarker}{\pgfqpoint{0.000000in}{-0.048611in}}{\pgfqpoint{0.000000in}{0.000000in}}{%
\pgfpathmoveto{\pgfqpoint{0.000000in}{0.000000in}}%
\pgfpathlineto{\pgfqpoint{0.000000in}{-0.048611in}}%
\pgfusepath{stroke,fill}%
}%
\begin{pgfscope}%
\pgfsys@transformshift{4.848876in}{0.502638in}%
\pgfsys@useobject{currentmarker}{}%
\end{pgfscope}%
\end{pgfscope}%
\begin{pgfscope}%
\definecolor{textcolor}{rgb}{0.000000,0.000000,0.000000}%
\pgfsetstrokecolor{textcolor}%
\pgfsetfillcolor{textcolor}%
\pgftext[x=4.848876in,y=0.405416in,,top]{\color{textcolor}\sffamily\fontsize{10.000000}{12.000000}\selectfont 1.0}%
\end{pgfscope}%
\begin{pgfscope}%
\definecolor{textcolor}{rgb}{0.000000,0.000000,0.000000}%
\pgfsetstrokecolor{textcolor}%
\pgfsetfillcolor{textcolor}%
\pgftext[x=2.805694in,y=0.226527in,,top]{\color{textcolor}\sffamily\fontsize{10.000000}{12.000000}\selectfont Fraction of edges used}%
\end{pgfscope}%
\begin{pgfscope}%
\pgfpathrectangle{\pgfqpoint{0.558194in}{0.502638in}}{\pgfqpoint{4.495000in}{1.848000in}}%
\pgfusepath{clip}%
\pgfsetrectcap%
\pgfsetroundjoin%
\pgfsetlinewidth{0.803000pt}%
\definecolor{currentstroke}{rgb}{0.690196,0.690196,0.690196}%
\pgfsetstrokecolor{currentstroke}%
\pgfsetdash{}{0pt}%
\pgfpathmoveto{\pgfqpoint{0.558194in}{0.567709in}}%
\pgfpathlineto{\pgfqpoint{5.053194in}{0.567709in}}%
\pgfusepath{stroke}%
\end{pgfscope}%
\begin{pgfscope}%
\pgfsetbuttcap%
\pgfsetroundjoin%
\definecolor{currentfill}{rgb}{0.000000,0.000000,0.000000}%
\pgfsetfillcolor{currentfill}%
\pgfsetlinewidth{0.803000pt}%
\definecolor{currentstroke}{rgb}{0.000000,0.000000,0.000000}%
\pgfsetstrokecolor{currentstroke}%
\pgfsetdash{}{0pt}%
\pgfsys@defobject{currentmarker}{\pgfqpoint{-0.048611in}{0.000000in}}{\pgfqpoint{0.000000in}{0.000000in}}{%
\pgfpathmoveto{\pgfqpoint{0.000000in}{0.000000in}}%
\pgfpathlineto{\pgfqpoint{-0.048611in}{0.000000in}}%
\pgfusepath{stroke,fill}%
}%
\begin{pgfscope}%
\pgfsys@transformshift{0.558194in}{0.567709in}%
\pgfsys@useobject{currentmarker}{}%
\end{pgfscope}%
\end{pgfscope}%
\begin{pgfscope}%
\definecolor{textcolor}{rgb}{0.000000,0.000000,0.000000}%
\pgfsetstrokecolor{textcolor}%
\pgfsetfillcolor{textcolor}%
\pgftext[x=0.283472in,y=0.519515in,left,base]{\color{textcolor}\sffamily\fontsize{10.000000}{12.000000}\selectfont 0.8}%
\end{pgfscope}%
\begin{pgfscope}%
\pgfpathrectangle{\pgfqpoint{0.558194in}{0.502638in}}{\pgfqpoint{4.495000in}{1.848000in}}%
\pgfusepath{clip}%
\pgfsetrectcap%
\pgfsetroundjoin%
\pgfsetlinewidth{0.803000pt}%
\definecolor{currentstroke}{rgb}{0.690196,0.690196,0.690196}%
\pgfsetstrokecolor{currentstroke}%
\pgfsetdash{}{0pt}%
\pgfpathmoveto{\pgfqpoint{0.558194in}{0.918954in}}%
\pgfpathlineto{\pgfqpoint{5.053194in}{0.918954in}}%
\pgfusepath{stroke}%
\end{pgfscope}%
\begin{pgfscope}%
\pgfsetbuttcap%
\pgfsetroundjoin%
\definecolor{currentfill}{rgb}{0.000000,0.000000,0.000000}%
\pgfsetfillcolor{currentfill}%
\pgfsetlinewidth{0.803000pt}%
\definecolor{currentstroke}{rgb}{0.000000,0.000000,0.000000}%
\pgfsetstrokecolor{currentstroke}%
\pgfsetdash{}{0pt}%
\pgfsys@defobject{currentmarker}{\pgfqpoint{-0.048611in}{0.000000in}}{\pgfqpoint{0.000000in}{0.000000in}}{%
\pgfpathmoveto{\pgfqpoint{0.000000in}{0.000000in}}%
\pgfpathlineto{\pgfqpoint{-0.048611in}{0.000000in}}%
\pgfusepath{stroke,fill}%
}%
\begin{pgfscope}%
\pgfsys@transformshift{0.558194in}{0.918954in}%
\pgfsys@useobject{currentmarker}{}%
\end{pgfscope}%
\end{pgfscope}%
\begin{pgfscope}%
\definecolor{textcolor}{rgb}{0.000000,0.000000,0.000000}%
\pgfsetstrokecolor{textcolor}%
\pgfsetfillcolor{textcolor}%
\pgftext[x=0.283472in,y=0.870760in,left,base]{\color{textcolor}\sffamily\fontsize{10.000000}{12.000000}\selectfont 0.9}%
\end{pgfscope}%
\begin{pgfscope}%
\pgfpathrectangle{\pgfqpoint{0.558194in}{0.502638in}}{\pgfqpoint{4.495000in}{1.848000in}}%
\pgfusepath{clip}%
\pgfsetrectcap%
\pgfsetroundjoin%
\pgfsetlinewidth{0.803000pt}%
\definecolor{currentstroke}{rgb}{0.690196,0.690196,0.690196}%
\pgfsetstrokecolor{currentstroke}%
\pgfsetdash{}{0pt}%
\pgfpathmoveto{\pgfqpoint{0.558194in}{1.270199in}}%
\pgfpathlineto{\pgfqpoint{5.053194in}{1.270199in}}%
\pgfusepath{stroke}%
\end{pgfscope}%
\begin{pgfscope}%
\pgfsetbuttcap%
\pgfsetroundjoin%
\definecolor{currentfill}{rgb}{0.000000,0.000000,0.000000}%
\pgfsetfillcolor{currentfill}%
\pgfsetlinewidth{0.803000pt}%
\definecolor{currentstroke}{rgb}{0.000000,0.000000,0.000000}%
\pgfsetstrokecolor{currentstroke}%
\pgfsetdash{}{0pt}%
\pgfsys@defobject{currentmarker}{\pgfqpoint{-0.048611in}{0.000000in}}{\pgfqpoint{0.000000in}{0.000000in}}{%
\pgfpathmoveto{\pgfqpoint{0.000000in}{0.000000in}}%
\pgfpathlineto{\pgfqpoint{-0.048611in}{0.000000in}}%
\pgfusepath{stroke,fill}%
}%
\begin{pgfscope}%
\pgfsys@transformshift{0.558194in}{1.270199in}%
\pgfsys@useobject{currentmarker}{}%
\end{pgfscope}%
\end{pgfscope}%
\begin{pgfscope}%
\definecolor{textcolor}{rgb}{0.000000,0.000000,0.000000}%
\pgfsetstrokecolor{textcolor}%
\pgfsetfillcolor{textcolor}%
\pgftext[x=0.283472in,y=1.222005in,left,base]{\color{textcolor}\sffamily\fontsize{10.000000}{12.000000}\selectfont 1.0}%
\end{pgfscope}%
\begin{pgfscope}%
\pgfpathrectangle{\pgfqpoint{0.558194in}{0.502638in}}{\pgfqpoint{4.495000in}{1.848000in}}%
\pgfusepath{clip}%
\pgfsetrectcap%
\pgfsetroundjoin%
\pgfsetlinewidth{0.803000pt}%
\definecolor{currentstroke}{rgb}{0.690196,0.690196,0.690196}%
\pgfsetstrokecolor{currentstroke}%
\pgfsetdash{}{0pt}%
\pgfpathmoveto{\pgfqpoint{0.558194in}{1.621444in}}%
\pgfpathlineto{\pgfqpoint{5.053194in}{1.621444in}}%
\pgfusepath{stroke}%
\end{pgfscope}%
\begin{pgfscope}%
\pgfsetbuttcap%
\pgfsetroundjoin%
\definecolor{currentfill}{rgb}{0.000000,0.000000,0.000000}%
\pgfsetfillcolor{currentfill}%
\pgfsetlinewidth{0.803000pt}%
\definecolor{currentstroke}{rgb}{0.000000,0.000000,0.000000}%
\pgfsetstrokecolor{currentstroke}%
\pgfsetdash{}{0pt}%
\pgfsys@defobject{currentmarker}{\pgfqpoint{-0.048611in}{0.000000in}}{\pgfqpoint{0.000000in}{0.000000in}}{%
\pgfpathmoveto{\pgfqpoint{0.000000in}{0.000000in}}%
\pgfpathlineto{\pgfqpoint{-0.048611in}{0.000000in}}%
\pgfusepath{stroke,fill}%
}%
\begin{pgfscope}%
\pgfsys@transformshift{0.558194in}{1.621444in}%
\pgfsys@useobject{currentmarker}{}%
\end{pgfscope}%
\end{pgfscope}%
\begin{pgfscope}%
\definecolor{textcolor}{rgb}{0.000000,0.000000,0.000000}%
\pgfsetstrokecolor{textcolor}%
\pgfsetfillcolor{textcolor}%
\pgftext[x=0.283472in,y=1.573249in,left,base]{\color{textcolor}\sffamily\fontsize{10.000000}{12.000000}\selectfont 1.1}%
\end{pgfscope}%
\begin{pgfscope}%
\pgfpathrectangle{\pgfqpoint{0.558194in}{0.502638in}}{\pgfqpoint{4.495000in}{1.848000in}}%
\pgfusepath{clip}%
\pgfsetrectcap%
\pgfsetroundjoin%
\pgfsetlinewidth{0.803000pt}%
\definecolor{currentstroke}{rgb}{0.690196,0.690196,0.690196}%
\pgfsetstrokecolor{currentstroke}%
\pgfsetdash{}{0pt}%
\pgfpathmoveto{\pgfqpoint{0.558194in}{1.972689in}}%
\pgfpathlineto{\pgfqpoint{5.053194in}{1.972689in}}%
\pgfusepath{stroke}%
\end{pgfscope}%
\begin{pgfscope}%
\pgfsetbuttcap%
\pgfsetroundjoin%
\definecolor{currentfill}{rgb}{0.000000,0.000000,0.000000}%
\pgfsetfillcolor{currentfill}%
\pgfsetlinewidth{0.803000pt}%
\definecolor{currentstroke}{rgb}{0.000000,0.000000,0.000000}%
\pgfsetstrokecolor{currentstroke}%
\pgfsetdash{}{0pt}%
\pgfsys@defobject{currentmarker}{\pgfqpoint{-0.048611in}{0.000000in}}{\pgfqpoint{0.000000in}{0.000000in}}{%
\pgfpathmoveto{\pgfqpoint{0.000000in}{0.000000in}}%
\pgfpathlineto{\pgfqpoint{-0.048611in}{0.000000in}}%
\pgfusepath{stroke,fill}%
}%
\begin{pgfscope}%
\pgfsys@transformshift{0.558194in}{1.972689in}%
\pgfsys@useobject{currentmarker}{}%
\end{pgfscope}%
\end{pgfscope}%
\begin{pgfscope}%
\definecolor{textcolor}{rgb}{0.000000,0.000000,0.000000}%
\pgfsetstrokecolor{textcolor}%
\pgfsetfillcolor{textcolor}%
\pgftext[x=0.283472in,y=1.924494in,left,base]{\color{textcolor}\sffamily\fontsize{10.000000}{12.000000}\selectfont 1.2}%
\end{pgfscope}%
\begin{pgfscope}%
\pgfpathrectangle{\pgfqpoint{0.558194in}{0.502638in}}{\pgfqpoint{4.495000in}{1.848000in}}%
\pgfusepath{clip}%
\pgfsetrectcap%
\pgfsetroundjoin%
\pgfsetlinewidth{0.803000pt}%
\definecolor{currentstroke}{rgb}{0.690196,0.690196,0.690196}%
\pgfsetstrokecolor{currentstroke}%
\pgfsetdash{}{0pt}%
\pgfpathmoveto{\pgfqpoint{0.558194in}{2.323934in}}%
\pgfpathlineto{\pgfqpoint{5.053194in}{2.323934in}}%
\pgfusepath{stroke}%
\end{pgfscope}%
\begin{pgfscope}%
\pgfsetbuttcap%
\pgfsetroundjoin%
\definecolor{currentfill}{rgb}{0.000000,0.000000,0.000000}%
\pgfsetfillcolor{currentfill}%
\pgfsetlinewidth{0.803000pt}%
\definecolor{currentstroke}{rgb}{0.000000,0.000000,0.000000}%
\pgfsetstrokecolor{currentstroke}%
\pgfsetdash{}{0pt}%
\pgfsys@defobject{currentmarker}{\pgfqpoint{-0.048611in}{0.000000in}}{\pgfqpoint{0.000000in}{0.000000in}}{%
\pgfpathmoveto{\pgfqpoint{0.000000in}{0.000000in}}%
\pgfpathlineto{\pgfqpoint{-0.048611in}{0.000000in}}%
\pgfusepath{stroke,fill}%
}%
\begin{pgfscope}%
\pgfsys@transformshift{0.558194in}{2.323934in}%
\pgfsys@useobject{currentmarker}{}%
\end{pgfscope}%
\end{pgfscope}%
\begin{pgfscope}%
\definecolor{textcolor}{rgb}{0.000000,0.000000,0.000000}%
\pgfsetstrokecolor{textcolor}%
\pgfsetfillcolor{textcolor}%
\pgftext[x=0.283472in,y=2.275739in,left,base]{\color{textcolor}\sffamily\fontsize{10.000000}{12.000000}\selectfont 1.3}%
\end{pgfscope}%
\begin{pgfscope}%
\definecolor{textcolor}{rgb}{0.000000,0.000000,0.000000}%
\pgfsetstrokecolor{textcolor}%
\pgfsetfillcolor{textcolor}%
\pgftext[x=0.227917in,y=1.426638in,,bottom,rotate=90.000000]{\color{textcolor}\sffamily\fontsize{10.000000}{12.000000}\selectfont Speedup over unstalled query}%
\end{pgfscope}%
\begin{pgfscope}%
\pgfpathrectangle{\pgfqpoint{0.558194in}{0.502638in}}{\pgfqpoint{4.495000in}{1.848000in}}%
\pgfusepath{clip}%
\pgfsetbuttcap%
\pgfsetroundjoin%
\pgfsetlinewidth{1.505625pt}%
\definecolor{currentstroke}{rgb}{0.000000,0.000000,1.000000}%
\pgfsetstrokecolor{currentstroke}%
\pgfsetdash{{5.550000pt}{2.400000pt}}{0.000000pt}%
\pgfpathmoveto{\pgfqpoint{0.762512in}{1.320021in}}%
\pgfpathlineto{\pgfqpoint{1.171149in}{1.195466in}}%
\pgfpathlineto{\pgfqpoint{1.579785in}{1.295110in}}%
\pgfpathlineto{\pgfqpoint{1.988422in}{1.344932in}}%
\pgfpathlineto{\pgfqpoint{2.397058in}{1.444576in}}%
\pgfpathlineto{\pgfqpoint{2.805694in}{1.569131in}}%
\pgfpathlineto{\pgfqpoint{3.214331in}{1.693686in}}%
\pgfpathlineto{\pgfqpoint{3.622967in}{1.818241in}}%
\pgfpathlineto{\pgfqpoint{4.031603in}{1.942796in}}%
\pgfpathlineto{\pgfqpoint{4.440240in}{2.092262in}}%
\pgfpathlineto{\pgfqpoint{4.848876in}{2.266638in}}%
\pgfusepath{stroke}%
\end{pgfscope}%
\begin{pgfscope}%
\pgfpathrectangle{\pgfqpoint{0.558194in}{0.502638in}}{\pgfqpoint{4.495000in}{1.848000in}}%
\pgfusepath{clip}%
\pgfsetbuttcap%
\pgfsetroundjoin%
\definecolor{currentfill}{rgb}{0.000000,0.000000,1.000000}%
\pgfsetfillcolor{currentfill}%
\pgfsetlinewidth{1.003750pt}%
\definecolor{currentstroke}{rgb}{0.000000,0.000000,1.000000}%
\pgfsetstrokecolor{currentstroke}%
\pgfsetdash{}{0pt}%
\pgfsys@defobject{currentmarker}{\pgfqpoint{-0.041667in}{-0.041667in}}{\pgfqpoint{0.041667in}{0.041667in}}{%
\pgfpathmoveto{\pgfqpoint{-0.041667in}{-0.041667in}}%
\pgfpathlineto{\pgfqpoint{0.041667in}{0.041667in}}%
\pgfpathmoveto{\pgfqpoint{-0.041667in}{0.041667in}}%
\pgfpathlineto{\pgfqpoint{0.041667in}{-0.041667in}}%
\pgfusepath{stroke,fill}%
}%
\begin{pgfscope}%
\pgfsys@transformshift{0.762512in}{1.320021in}%
\pgfsys@useobject{currentmarker}{}%
\end{pgfscope}%
\begin{pgfscope}%
\pgfsys@transformshift{1.171149in}{1.195466in}%
\pgfsys@useobject{currentmarker}{}%
\end{pgfscope}%
\begin{pgfscope}%
\pgfsys@transformshift{1.579785in}{1.295110in}%
\pgfsys@useobject{currentmarker}{}%
\end{pgfscope}%
\begin{pgfscope}%
\pgfsys@transformshift{1.988422in}{1.344932in}%
\pgfsys@useobject{currentmarker}{}%
\end{pgfscope}%
\begin{pgfscope}%
\pgfsys@transformshift{2.397058in}{1.444576in}%
\pgfsys@useobject{currentmarker}{}%
\end{pgfscope}%
\begin{pgfscope}%
\pgfsys@transformshift{2.805694in}{1.569131in}%
\pgfsys@useobject{currentmarker}{}%
\end{pgfscope}%
\begin{pgfscope}%
\pgfsys@transformshift{3.214331in}{1.693686in}%
\pgfsys@useobject{currentmarker}{}%
\end{pgfscope}%
\begin{pgfscope}%
\pgfsys@transformshift{3.622967in}{1.818241in}%
\pgfsys@useobject{currentmarker}{}%
\end{pgfscope}%
\begin{pgfscope}%
\pgfsys@transformshift{4.031603in}{1.942796in}%
\pgfsys@useobject{currentmarker}{}%
\end{pgfscope}%
\begin{pgfscope}%
\pgfsys@transformshift{4.440240in}{2.092262in}%
\pgfsys@useobject{currentmarker}{}%
\end{pgfscope}%
\begin{pgfscope}%
\pgfsys@transformshift{4.848876in}{2.266638in}%
\pgfsys@useobject{currentmarker}{}%
\end{pgfscope}%
\end{pgfscope}%
\begin{pgfscope}%
\pgfpathrectangle{\pgfqpoint{0.558194in}{0.502638in}}{\pgfqpoint{4.495000in}{1.848000in}}%
\pgfusepath{clip}%
\pgfsetrectcap%
\pgfsetroundjoin%
\pgfsetlinewidth{1.505625pt}%
\definecolor{currentstroke}{rgb}{0.000000,0.000000,1.000000}%
\pgfsetstrokecolor{currentstroke}%
\pgfsetdash{}{0pt}%
\pgfpathmoveto{\pgfqpoint{0.762512in}{1.341158in}}%
\pgfpathlineto{\pgfqpoint{1.171149in}{1.146021in}}%
\pgfpathlineto{\pgfqpoint{1.579785in}{1.216980in}}%
\pgfpathlineto{\pgfqpoint{1.988422in}{1.287939in}}%
\pgfpathlineto{\pgfqpoint{2.397058in}{1.376637in}}%
\pgfpathlineto{\pgfqpoint{2.805694in}{1.500814in}}%
\pgfpathlineto{\pgfqpoint{3.214331in}{1.624992in}}%
\pgfpathlineto{\pgfqpoint{3.622967in}{1.749169in}}%
\pgfpathlineto{\pgfqpoint{4.031603in}{1.908826in}}%
\pgfpathlineto{\pgfqpoint{4.440240in}{2.033004in}}%
\pgfpathlineto{\pgfqpoint{4.848876in}{2.192660in}}%
\pgfusepath{stroke}%
\end{pgfscope}%
\begin{pgfscope}%
\pgfpathrectangle{\pgfqpoint{0.558194in}{0.502638in}}{\pgfqpoint{4.495000in}{1.848000in}}%
\pgfusepath{clip}%
\pgfsetbuttcap%
\pgfsetroundjoin%
\definecolor{currentfill}{rgb}{0.000000,0.000000,1.000000}%
\pgfsetfillcolor{currentfill}%
\pgfsetlinewidth{1.003750pt}%
\definecolor{currentstroke}{rgb}{0.000000,0.000000,1.000000}%
\pgfsetstrokecolor{currentstroke}%
\pgfsetdash{}{0pt}%
\pgfsys@defobject{currentmarker}{\pgfqpoint{-0.041667in}{-0.041667in}}{\pgfqpoint{0.041667in}{0.041667in}}{%
\pgfpathmoveto{\pgfqpoint{-0.041667in}{-0.041667in}}%
\pgfpathlineto{\pgfqpoint{0.041667in}{0.041667in}}%
\pgfpathmoveto{\pgfqpoint{-0.041667in}{0.041667in}}%
\pgfpathlineto{\pgfqpoint{0.041667in}{-0.041667in}}%
\pgfusepath{stroke,fill}%
}%
\begin{pgfscope}%
\pgfsys@transformshift{0.762512in}{1.341158in}%
\pgfsys@useobject{currentmarker}{}%
\end{pgfscope}%
\begin{pgfscope}%
\pgfsys@transformshift{1.171149in}{1.146021in}%
\pgfsys@useobject{currentmarker}{}%
\end{pgfscope}%
\begin{pgfscope}%
\pgfsys@transformshift{1.579785in}{1.216980in}%
\pgfsys@useobject{currentmarker}{}%
\end{pgfscope}%
\begin{pgfscope}%
\pgfsys@transformshift{1.988422in}{1.287939in}%
\pgfsys@useobject{currentmarker}{}%
\end{pgfscope}%
\begin{pgfscope}%
\pgfsys@transformshift{2.397058in}{1.376637in}%
\pgfsys@useobject{currentmarker}{}%
\end{pgfscope}%
\begin{pgfscope}%
\pgfsys@transformshift{2.805694in}{1.500814in}%
\pgfsys@useobject{currentmarker}{}%
\end{pgfscope}%
\begin{pgfscope}%
\pgfsys@transformshift{3.214331in}{1.624992in}%
\pgfsys@useobject{currentmarker}{}%
\end{pgfscope}%
\begin{pgfscope}%
\pgfsys@transformshift{3.622967in}{1.749169in}%
\pgfsys@useobject{currentmarker}{}%
\end{pgfscope}%
\begin{pgfscope}%
\pgfsys@transformshift{4.031603in}{1.908826in}%
\pgfsys@useobject{currentmarker}{}%
\end{pgfscope}%
\begin{pgfscope}%
\pgfsys@transformshift{4.440240in}{2.033004in}%
\pgfsys@useobject{currentmarker}{}%
\end{pgfscope}%
\begin{pgfscope}%
\pgfsys@transformshift{4.848876in}{2.192660in}%
\pgfsys@useobject{currentmarker}{}%
\end{pgfscope}%
\end{pgfscope}%
\begin{pgfscope}%
\pgfpathrectangle{\pgfqpoint{0.558194in}{0.502638in}}{\pgfqpoint{4.495000in}{1.848000in}}%
\pgfusepath{clip}%
\pgfsetbuttcap%
\pgfsetroundjoin%
\pgfsetlinewidth{1.505625pt}%
\definecolor{currentstroke}{rgb}{1.000000,0.647059,0.000000}%
\pgfsetstrokecolor{currentstroke}%
\pgfsetdash{{5.550000pt}{2.400000pt}}{0.000000pt}%
\pgfpathmoveto{\pgfqpoint{0.762512in}{1.343813in}}%
\pgfpathlineto{\pgfqpoint{1.171149in}{0.849546in}}%
\pgfpathlineto{\pgfqpoint{1.579785in}{0.654995in}}%
\pgfpathlineto{\pgfqpoint{1.988422in}{0.586638in}}%
\pgfpathlineto{\pgfqpoint{2.397058in}{0.591897in}}%
\pgfpathlineto{\pgfqpoint{2.805694in}{0.644478in}}%
\pgfpathlineto{\pgfqpoint{3.214331in}{0.723351in}}%
\pgfpathlineto{\pgfqpoint{3.622967in}{0.817997in}}%
\pgfpathlineto{\pgfqpoint{4.031603in}{0.938935in}}%
\pgfpathlineto{\pgfqpoint{4.440240in}{1.065131in}}%
\pgfpathlineto{\pgfqpoint{4.848876in}{1.217617in}}%
\pgfusepath{stroke}%
\end{pgfscope}%
\begin{pgfscope}%
\pgfpathrectangle{\pgfqpoint{0.558194in}{0.502638in}}{\pgfqpoint{4.495000in}{1.848000in}}%
\pgfusepath{clip}%
\pgfsetbuttcap%
\pgfsetroundjoin%
\definecolor{currentfill}{rgb}{1.000000,0.647059,0.000000}%
\pgfsetfillcolor{currentfill}%
\pgfsetlinewidth{1.003750pt}%
\definecolor{currentstroke}{rgb}{1.000000,0.647059,0.000000}%
\pgfsetstrokecolor{currentstroke}%
\pgfsetdash{}{0pt}%
\pgfsys@defobject{currentmarker}{\pgfqpoint{-0.041667in}{-0.041667in}}{\pgfqpoint{0.041667in}{0.041667in}}{%
\pgfpathmoveto{\pgfqpoint{-0.041667in}{-0.041667in}}%
\pgfpathlineto{\pgfqpoint{0.041667in}{0.041667in}}%
\pgfpathmoveto{\pgfqpoint{-0.041667in}{0.041667in}}%
\pgfpathlineto{\pgfqpoint{0.041667in}{-0.041667in}}%
\pgfusepath{stroke,fill}%
}%
\begin{pgfscope}%
\pgfsys@transformshift{0.762512in}{1.343813in}%
\pgfsys@useobject{currentmarker}{}%
\end{pgfscope}%
\begin{pgfscope}%
\pgfsys@transformshift{1.171149in}{0.849546in}%
\pgfsys@useobject{currentmarker}{}%
\end{pgfscope}%
\begin{pgfscope}%
\pgfsys@transformshift{1.579785in}{0.654995in}%
\pgfsys@useobject{currentmarker}{}%
\end{pgfscope}%
\begin{pgfscope}%
\pgfsys@transformshift{1.988422in}{0.586638in}%
\pgfsys@useobject{currentmarker}{}%
\end{pgfscope}%
\begin{pgfscope}%
\pgfsys@transformshift{2.397058in}{0.591897in}%
\pgfsys@useobject{currentmarker}{}%
\end{pgfscope}%
\begin{pgfscope}%
\pgfsys@transformshift{2.805694in}{0.644478in}%
\pgfsys@useobject{currentmarker}{}%
\end{pgfscope}%
\begin{pgfscope}%
\pgfsys@transformshift{3.214331in}{0.723351in}%
\pgfsys@useobject{currentmarker}{}%
\end{pgfscope}%
\begin{pgfscope}%
\pgfsys@transformshift{3.622967in}{0.817997in}%
\pgfsys@useobject{currentmarker}{}%
\end{pgfscope}%
\begin{pgfscope}%
\pgfsys@transformshift{4.031603in}{0.938935in}%
\pgfsys@useobject{currentmarker}{}%
\end{pgfscope}%
\begin{pgfscope}%
\pgfsys@transformshift{4.440240in}{1.065131in}%
\pgfsys@useobject{currentmarker}{}%
\end{pgfscope}%
\begin{pgfscope}%
\pgfsys@transformshift{4.848876in}{1.217617in}%
\pgfsys@useobject{currentmarker}{}%
\end{pgfscope}%
\end{pgfscope}%
\begin{pgfscope}%
\pgfpathrectangle{\pgfqpoint{0.558194in}{0.502638in}}{\pgfqpoint{4.495000in}{1.848000in}}%
\pgfusepath{clip}%
\pgfsetrectcap%
\pgfsetroundjoin%
\pgfsetlinewidth{1.505625pt}%
\definecolor{currentstroke}{rgb}{1.000000,0.647059,0.000000}%
\pgfsetstrokecolor{currentstroke}%
\pgfsetdash{}{0pt}%
\pgfpathmoveto{\pgfqpoint{0.762512in}{1.368087in}}%
\pgfpathlineto{\pgfqpoint{1.171149in}{1.005326in}}%
\pgfpathlineto{\pgfqpoint{1.579785in}{0.895922in}}%
\pgfpathlineto{\pgfqpoint{1.988422in}{0.826824in}}%
\pgfpathlineto{\pgfqpoint{2.397058in}{0.832582in}}%
\pgfpathlineto{\pgfqpoint{2.805694in}{0.895922in}}%
\pgfpathlineto{\pgfqpoint{3.214331in}{0.976535in}}%
\pgfpathlineto{\pgfqpoint{3.622967in}{1.085939in}}%
\pgfpathlineto{\pgfqpoint{4.031603in}{1.218376in}}%
\pgfpathlineto{\pgfqpoint{4.440240in}{1.345054in}}%
\pgfpathlineto{\pgfqpoint{4.848876in}{1.489007in}}%
\pgfusepath{stroke}%
\end{pgfscope}%
\begin{pgfscope}%
\pgfpathrectangle{\pgfqpoint{0.558194in}{0.502638in}}{\pgfqpoint{4.495000in}{1.848000in}}%
\pgfusepath{clip}%
\pgfsetbuttcap%
\pgfsetroundjoin%
\definecolor{currentfill}{rgb}{1.000000,0.647059,0.000000}%
\pgfsetfillcolor{currentfill}%
\pgfsetlinewidth{1.003750pt}%
\definecolor{currentstroke}{rgb}{1.000000,0.647059,0.000000}%
\pgfsetstrokecolor{currentstroke}%
\pgfsetdash{}{0pt}%
\pgfsys@defobject{currentmarker}{\pgfqpoint{-0.041667in}{-0.041667in}}{\pgfqpoint{0.041667in}{0.041667in}}{%
\pgfpathmoveto{\pgfqpoint{-0.041667in}{-0.041667in}}%
\pgfpathlineto{\pgfqpoint{0.041667in}{0.041667in}}%
\pgfpathmoveto{\pgfqpoint{-0.041667in}{0.041667in}}%
\pgfpathlineto{\pgfqpoint{0.041667in}{-0.041667in}}%
\pgfusepath{stroke,fill}%
}%
\begin{pgfscope}%
\pgfsys@transformshift{0.762512in}{1.368087in}%
\pgfsys@useobject{currentmarker}{}%
\end{pgfscope}%
\begin{pgfscope}%
\pgfsys@transformshift{1.171149in}{1.005326in}%
\pgfsys@useobject{currentmarker}{}%
\end{pgfscope}%
\begin{pgfscope}%
\pgfsys@transformshift{1.579785in}{0.895922in}%
\pgfsys@useobject{currentmarker}{}%
\end{pgfscope}%
\begin{pgfscope}%
\pgfsys@transformshift{1.988422in}{0.826824in}%
\pgfsys@useobject{currentmarker}{}%
\end{pgfscope}%
\begin{pgfscope}%
\pgfsys@transformshift{2.397058in}{0.832582in}%
\pgfsys@useobject{currentmarker}{}%
\end{pgfscope}%
\begin{pgfscope}%
\pgfsys@transformshift{2.805694in}{0.895922in}%
\pgfsys@useobject{currentmarker}{}%
\end{pgfscope}%
\begin{pgfscope}%
\pgfsys@transformshift{3.214331in}{0.976535in}%
\pgfsys@useobject{currentmarker}{}%
\end{pgfscope}%
\begin{pgfscope}%
\pgfsys@transformshift{3.622967in}{1.085939in}%
\pgfsys@useobject{currentmarker}{}%
\end{pgfscope}%
\begin{pgfscope}%
\pgfsys@transformshift{4.031603in}{1.218376in}%
\pgfsys@useobject{currentmarker}{}%
\end{pgfscope}%
\begin{pgfscope}%
\pgfsys@transformshift{4.440240in}{1.345054in}%
\pgfsys@useobject{currentmarker}{}%
\end{pgfscope}%
\begin{pgfscope}%
\pgfsys@transformshift{4.848876in}{1.489007in}%
\pgfsys@useobject{currentmarker}{}%
\end{pgfscope}%
\end{pgfscope}%
\begin{pgfscope}%
\pgfsetrectcap%
\pgfsetmiterjoin%
\pgfsetlinewidth{0.803000pt}%
\definecolor{currentstroke}{rgb}{0.000000,0.000000,0.000000}%
\pgfsetstrokecolor{currentstroke}%
\pgfsetdash{}{0pt}%
\pgfpathmoveto{\pgfqpoint{0.558194in}{0.502638in}}%
\pgfpathlineto{\pgfqpoint{0.558194in}{2.350638in}}%
\pgfusepath{stroke}%
\end{pgfscope}%
\begin{pgfscope}%
\pgfsetrectcap%
\pgfsetmiterjoin%
\pgfsetlinewidth{0.803000pt}%
\definecolor{currentstroke}{rgb}{0.000000,0.000000,0.000000}%
\pgfsetstrokecolor{currentstroke}%
\pgfsetdash{}{0pt}%
\pgfpathmoveto{\pgfqpoint{5.053194in}{0.502638in}}%
\pgfpathlineto{\pgfqpoint{5.053194in}{2.350638in}}%
\pgfusepath{stroke}%
\end{pgfscope}%
\begin{pgfscope}%
\pgfsetrectcap%
\pgfsetmiterjoin%
\pgfsetlinewidth{0.803000pt}%
\definecolor{currentstroke}{rgb}{0.000000,0.000000,0.000000}%
\pgfsetstrokecolor{currentstroke}%
\pgfsetdash{}{0pt}%
\pgfpathmoveto{\pgfqpoint{0.558194in}{0.502638in}}%
\pgfpathlineto{\pgfqpoint{5.053194in}{0.502638in}}%
\pgfusepath{stroke}%
\end{pgfscope}%
\begin{pgfscope}%
\pgfsetrectcap%
\pgfsetmiterjoin%
\pgfsetlinewidth{0.803000pt}%
\definecolor{currentstroke}{rgb}{0.000000,0.000000,0.000000}%
\pgfsetstrokecolor{currentstroke}%
\pgfsetdash{}{0pt}%
\pgfpathmoveto{\pgfqpoint{0.558194in}{2.350638in}}%
\pgfpathlineto{\pgfqpoint{5.053194in}{2.350638in}}%
\pgfusepath{stroke}%
\end{pgfscope}%
\begin{pgfscope}%
\pgfsetbuttcap%
\pgfsetmiterjoin%
\definecolor{currentfill}{rgb}{1.000000,1.000000,1.000000}%
\pgfsetfillcolor{currentfill}%
\pgfsetfillopacity{0.800000}%
\pgfsetlinewidth{1.003750pt}%
\definecolor{currentstroke}{rgb}{0.800000,0.800000,0.800000}%
\pgfsetstrokecolor{currentstroke}%
\pgfsetstrokeopacity{0.800000}%
\pgfsetdash{}{0pt}%
\pgfpathmoveto{\pgfqpoint{0.655417in}{1.465083in}}%
\pgfpathlineto{\pgfqpoint{2.219722in}{1.465083in}}%
\pgfpathquadraticcurveto{\pgfqpoint{2.247500in}{1.465083in}}{\pgfqpoint{2.247500in}{1.492861in}}%
\pgfpathlineto{\pgfqpoint{2.247500in}{2.253416in}}%
\pgfpathquadraticcurveto{\pgfqpoint{2.247500in}{2.281194in}}{\pgfqpoint{2.219722in}{2.281194in}}%
\pgfpathlineto{\pgfqpoint{0.655417in}{2.281194in}}%
\pgfpathquadraticcurveto{\pgfqpoint{0.627639in}{2.281194in}}{\pgfqpoint{0.627639in}{2.253416in}}%
\pgfpathlineto{\pgfqpoint{0.627639in}{1.492861in}}%
\pgfpathquadraticcurveto{\pgfqpoint{0.627639in}{1.465083in}}{\pgfqpoint{0.655417in}{1.465083in}}%
\pgfpathclose%
\pgfusepath{stroke,fill}%
\end{pgfscope}%
\begin{pgfscope}%
\pgfsetbuttcap%
\pgfsetroundjoin%
\pgfsetlinewidth{1.505625pt}%
\definecolor{currentstroke}{rgb}{0.000000,0.000000,1.000000}%
\pgfsetstrokecolor{currentstroke}%
\pgfsetdash{{5.550000pt}{2.400000pt}}{0.000000pt}%
\pgfpathmoveto{\pgfqpoint{0.683194in}{2.177027in}}%
\pgfpathlineto{\pgfqpoint{0.960972in}{2.177027in}}%
\pgfusepath{stroke}%
\end{pgfscope}%
\begin{pgfscope}%
\pgfsetbuttcap%
\pgfsetroundjoin%
\definecolor{currentfill}{rgb}{0.000000,0.000000,1.000000}%
\pgfsetfillcolor{currentfill}%
\pgfsetlinewidth{1.003750pt}%
\definecolor{currentstroke}{rgb}{0.000000,0.000000,1.000000}%
\pgfsetstrokecolor{currentstroke}%
\pgfsetdash{}{0pt}%
\pgfsys@defobject{currentmarker}{\pgfqpoint{-0.041667in}{-0.041667in}}{\pgfqpoint{0.041667in}{0.041667in}}{%
\pgfpathmoveto{\pgfqpoint{-0.041667in}{-0.041667in}}%
\pgfpathlineto{\pgfqpoint{0.041667in}{0.041667in}}%
\pgfpathmoveto{\pgfqpoint{-0.041667in}{0.041667in}}%
\pgfpathlineto{\pgfqpoint{0.041667in}{-0.041667in}}%
\pgfusepath{stroke,fill}%
}%
\begin{pgfscope}%
\pgfsys@transformshift{0.822083in}{2.177027in}%
\pgfsys@useobject{currentmarker}{}%
\end{pgfscope}%
\end{pgfscope}%
\begin{pgfscope}%
\definecolor{textcolor}{rgb}{0.000000,0.000000,0.000000}%
\pgfsetstrokecolor{textcolor}%
\pgfsetfillcolor{textcolor}%
\pgftext[x=1.072083in,y=2.128416in,left,base]{\color{textcolor}\sffamily\fontsize{10.000000}{12.000000}\selectfont travel time}%
\end{pgfscope}%
\begin{pgfscope}%
\pgfsetrectcap%
\pgfsetroundjoin%
\pgfsetlinewidth{1.505625pt}%
\definecolor{currentstroke}{rgb}{0.000000,0.000000,1.000000}%
\pgfsetstrokecolor{currentstroke}%
\pgfsetdash{}{0pt}%
\pgfpathmoveto{\pgfqpoint{0.683194in}{1.983416in}}%
\pgfpathlineto{\pgfqpoint{0.960972in}{1.983416in}}%
\pgfusepath{stroke}%
\end{pgfscope}%
\begin{pgfscope}%
\pgfsetbuttcap%
\pgfsetroundjoin%
\definecolor{currentfill}{rgb}{0.000000,0.000000,1.000000}%
\pgfsetfillcolor{currentfill}%
\pgfsetlinewidth{1.003750pt}%
\definecolor{currentstroke}{rgb}{0.000000,0.000000,1.000000}%
\pgfsetstrokecolor{currentstroke}%
\pgfsetdash{}{0pt}%
\pgfsys@defobject{currentmarker}{\pgfqpoint{-0.041667in}{-0.041667in}}{\pgfqpoint{0.041667in}{0.041667in}}{%
\pgfpathmoveto{\pgfqpoint{-0.041667in}{-0.041667in}}%
\pgfpathlineto{\pgfqpoint{0.041667in}{0.041667in}}%
\pgfpathmoveto{\pgfqpoint{-0.041667in}{0.041667in}}%
\pgfpathlineto{\pgfqpoint{0.041667in}{-0.041667in}}%
\pgfusepath{stroke,fill}%
}%
\begin{pgfscope}%
\pgfsys@transformshift{0.822083in}{1.983416in}%
\pgfsys@useobject{currentmarker}{}%
\end{pgfscope}%
\end{pgfscope}%
\begin{pgfscope}%
\definecolor{textcolor}{rgb}{0.000000,0.000000,0.000000}%
\pgfsetstrokecolor{textcolor}%
\pgfsetfillcolor{textcolor}%
\pgftext[x=1.072083in,y=1.934805in,left,base]{\color{textcolor}\sffamily\fontsize{10.000000}{12.000000}\selectfont travel time + turns}%
\end{pgfscope}%
\begin{pgfscope}%
\pgfsetbuttcap%
\pgfsetroundjoin%
\pgfsetlinewidth{1.505625pt}%
\definecolor{currentstroke}{rgb}{1.000000,0.647059,0.000000}%
\pgfsetstrokecolor{currentstroke}%
\pgfsetdash{{5.550000pt}{2.400000pt}}{0.000000pt}%
\pgfpathmoveto{\pgfqpoint{0.683194in}{1.789805in}}%
\pgfpathlineto{\pgfqpoint{0.960972in}{1.789805in}}%
\pgfusepath{stroke}%
\end{pgfscope}%
\begin{pgfscope}%
\pgfsetbuttcap%
\pgfsetroundjoin%
\definecolor{currentfill}{rgb}{1.000000,0.647059,0.000000}%
\pgfsetfillcolor{currentfill}%
\pgfsetlinewidth{1.003750pt}%
\definecolor{currentstroke}{rgb}{1.000000,0.647059,0.000000}%
\pgfsetstrokecolor{currentstroke}%
\pgfsetdash{}{0pt}%
\pgfsys@defobject{currentmarker}{\pgfqpoint{-0.041667in}{-0.041667in}}{\pgfqpoint{0.041667in}{0.041667in}}{%
\pgfpathmoveto{\pgfqpoint{-0.041667in}{-0.041667in}}%
\pgfpathlineto{\pgfqpoint{0.041667in}{0.041667in}}%
\pgfpathmoveto{\pgfqpoint{-0.041667in}{0.041667in}}%
\pgfpathlineto{\pgfqpoint{0.041667in}{-0.041667in}}%
\pgfusepath{stroke,fill}%
}%
\begin{pgfscope}%
\pgfsys@transformshift{0.822083in}{1.789805in}%
\pgfsys@useobject{currentmarker}{}%
\end{pgfscope}%
\end{pgfscope}%
\begin{pgfscope}%
\definecolor{textcolor}{rgb}{0.000000,0.000000,0.000000}%
\pgfsetstrokecolor{textcolor}%
\pgfsetfillcolor{textcolor}%
\pgftext[x=1.072083in,y=1.741194in,left,base]{\color{textcolor}\sffamily\fontsize{10.000000}{12.000000}\selectfont distance}%
\end{pgfscope}%
\begin{pgfscope}%
\pgfsetrectcap%
\pgfsetroundjoin%
\pgfsetlinewidth{1.505625pt}%
\definecolor{currentstroke}{rgb}{1.000000,0.647059,0.000000}%
\pgfsetstrokecolor{currentstroke}%
\pgfsetdash{}{0pt}%
\pgfpathmoveto{\pgfqpoint{0.683194in}{1.596194in}}%
\pgfpathlineto{\pgfqpoint{0.960972in}{1.596194in}}%
\pgfusepath{stroke}%
\end{pgfscope}%
\begin{pgfscope}%
\pgfsetbuttcap%
\pgfsetroundjoin%
\definecolor{currentfill}{rgb}{1.000000,0.647059,0.000000}%
\pgfsetfillcolor{currentfill}%
\pgfsetlinewidth{1.003750pt}%
\definecolor{currentstroke}{rgb}{1.000000,0.647059,0.000000}%
\pgfsetstrokecolor{currentstroke}%
\pgfsetdash{}{0pt}%
\pgfsys@defobject{currentmarker}{\pgfqpoint{-0.041667in}{-0.041667in}}{\pgfqpoint{0.041667in}{0.041667in}}{%
\pgfpathmoveto{\pgfqpoint{-0.041667in}{-0.041667in}}%
\pgfpathlineto{\pgfqpoint{0.041667in}{0.041667in}}%
\pgfpathmoveto{\pgfqpoint{-0.041667in}{0.041667in}}%
\pgfpathlineto{\pgfqpoint{0.041667in}{-0.041667in}}%
\pgfusepath{stroke,fill}%
}%
\begin{pgfscope}%
\pgfsys@transformshift{0.822083in}{1.596194in}%
\pgfsys@useobject{currentmarker}{}%
\end{pgfscope}%
\end{pgfscope}%
\begin{pgfscope}%
\definecolor{textcolor}{rgb}{0.000000,0.000000,0.000000}%
\pgfsetstrokecolor{textcolor}%
\pgfsetfillcolor{textcolor}%
\pgftext[x=1.072083in,y=1.547583in,left,base]{\color{textcolor}\sffamily\fontsize{10.000000}{12.000000}\selectfont distance + turns}%
\end{pgfscope}%
\end{pgfpicture}%
\makeatother%
\endgroup%